\documentclass[12pt]{iopart}
\usepackage{epsfig,graphicx,multicol,bbm,mathrsfs,amssymb,dcolumn,bm}

\begin{document}

\title[Three-dimensional track reconstruction for directional Dark Matter detection]{
Three-dimensional track reconstruction for directional Dark Matter detection
}
\author{J. Billard, F. Mayet and D. Santos}
\address{Laboratoire de Physique Subatomique et de Cosmologie, Universit\'e Joseph Fourier Grenoble 1,
  CNRS/IN2P3, Institut Nationale Polytechnique de Grenoble, 53 rue des Martyrs, Grenoble, France }
\ead{billard@lpsc.in2p3.fr}

\begin{abstract}
Directional detection of Dark Matter is a promising search strategy.
However, to perform such detection,   a given set of parameters has to be retrieved from the recoiling tracks : direction, sense and position in the detector volume.
In order to optimize the track reconstruction and to fully exploit the data of forthcoming directional detectors, we present  
a likelihood method dedicated to 3D track reconstruction. This new analysis method is 
applied to the MIMAC detector. It requires a 
full simulation of track measurements in order to compare real tracks to simulated ones.  We conclude that a good spatial resolution can be achieved, {\it
i.e.}  sub-mm in the anode plane and cm along the drift axis. This opens the possibility to perform a fiducialization of 
directional detectors. The angular resolution is shown to range between 20$^\circ$ to 80$^\circ$, depending on the recoil energy, which is
however enough to achieve a high significance discovery of Dark Matter. On the contrary, 
we show that sense recognition capability of directional detectors depends strongly on the recoil energy and the drift distance, 
with small efficiency values (50\%-70\%). We suggest not to consider this
information either for exclusion or discovery of Dark Matter for recoils below 100 keV and then 
to focus on axial directional data.

\end{abstract}

\maketitle


Directional detection of galactic Dark Matter offers
a unique opportunity to identify Weakly Interacting Massive Particle (WIMP) events as
such~\cite{spergel}. Recent studies have shown that a low exposure directional detector could lead either to  
a high significance discovery of galactic Dark Matter \cite{billard.disco,billard.profile,billard.ident,green.disco,albornoz} or to a conclusive exclusion \cite{billard.exclusion}, 
depending on the value of the unknown  WIMP-nucleon cross section.\\ 
Directional detection  requires  the simultaneous measurement of the recoil energy ($E_R$) and the direction of the 3D track ($\Omega_R$) of 
low energy recoils, thus  allowing to evaluate the double-differential spectrum $\mathrm{d}^2R/\mathrm{d}E_R\mathrm{d}\Omega_R$ 
down to the energy threshold. This can be achieved with low pressure Time Projection Chamber (TPC) detectors  and 
there is a worldwide effort toward the development of a large TPC devoted to directional detection \cite{white}. All current 
projects \cite{dmtpc,drift,d3,mimac,newage} face common challenges amongst which the 3D reconstruction 
of low energy tracks ($\mathcal{O}(10)$ keV) is the major one as it includes various experimental issues such as sense recognition, angular and energy
resolutions and energy threshold. Their effect on the discovery potential of forthcoming directional 
detectors has been fully studied in \cite{billard.profile}. The aim of this paper is twofold.\\ 
First a dedicated 3D reconstruction data analysis is proposed. The goal is to retrieve, for each track, the initial 
recoil direction ($\theta, \phi$) and the vertex (X, Y and Z) of the elastic scatterring interaction. 
Difficulties come from the fact that the recoil energy is low ($E_r \lesssim 100$ keV) and the track length is small ($\lesssim 10$ mm).
For instance, in the case of a Fluorine target at 50 mbar, the mean track length of a 10 keV recoil is of the order of 300 $\mu m$ and about 3 mm at 100 keV. 
Moreover, low energy recoil in low pressure TPC will encounter a rather large angular dispersion which implies an intrinsic angular resolution. 
Then, the electron drift within the TPC is characterized by a transverse and longitudinal diffusion, which results in an embedding of the initial track within a rather large 
 envelope. 
Finally, even a high performance readout induces a pixelization of the track. These effects highlight the need to go beyond 
a straightforward track reconstruction algorithm, {\it i.e.} a three dimensional linear fit. 
The method proposed in this paper is of general interest for directional projects, although a few details should be modified to  follow particular readout strategies.\\
Second goal of this paper is to evaluate the expected performance of the MIMAC project \cite{mimac}, 
in terms of angular resolution, fiducialization and sense recognition capability. This study is based on
simulations and is closely related to the discovery potential, 
as outlined in~\cite{billard.profile}.\\

The paper is organised as follows, we first introduce the MIMAC track measurement strategy and 
the different observables given by the MIMAC readouts (sec.~\ref{sec:det}). In section~\ref{sec:simu}, 
we give a comprehensive overview of various systematics of track measurement before presenting the simulation used to generate mock data. 
Then, we introduce a new analysis method in section \ref{sec:method}, based on a full likelihood approach combined with a multivariate 
analysis. Section \ref{sec:results} presents the expected MIMAC performance (spatial and angular resolutions, 
fiducialisation, senses recognition capability)  as a function of the energy and of the $Z$ coordinate (perpendicular to the anode plane).

\section{Directional detection framework and MIMAC characteristics}
\label{sec:det}
\begin{figure*}[t]
\begin{center}
\includegraphics[scale=0.4,angle=0]{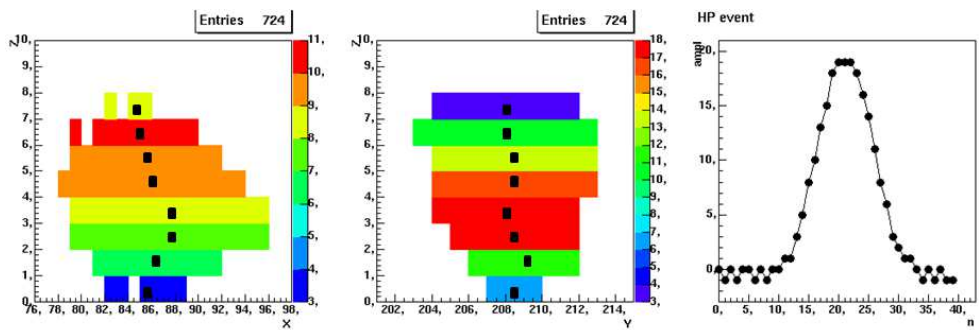}
\caption{Measurement of a 3 dimensional track corresponding to a Fluorine recoil candidate at 50 keVee in a gas mixture of 70\% $CF_4$ + 30\% $CHF_3$ at 50 mbar. The left and
center panel represent the projection of the track on the (X,Z) and (Y,Z) plane. The color of the different time slice (along the Z axis) refers to the number of activated
strips along the third dimension. The black boxes represent the position of the center of gravity of each time slice. Right panel represent the derivative of the charge
integrator response. We remind that both the anode and the charge preamplifier are sampled at a frequency of 50 MHz.}  
\label{fig:Observable}
\end{center}
\end{figure*}

Several Dark Matter directional detectors are being developed and/or operated \cite{white,dmtpc,drift,d3,mimac,newage}. They all  
face common challenges amongst which the 3D reconstruction of low energy tracks ($\mathcal{O}(10)$ keV) is the major one. 
Indeed,  a directional detector should allow to evaluate  the double-differential spectrum 
$\mathrm{d}^2R/\mathrm{d}E_R\mathrm{d}\Omega_R$ down to the recoil energy threshold. We emphasize that it is the lowest energy at
which one can retrieve both the initial direction and the energy of the
recoiling nucleus, taking into account the ionization quenching factor. This makes it even
more challenging for directional detection. In the following, we present the MIMAC track reconstruction strategy 
(sec.~\ref{sec:mimac}), discuss the observables and outline the need to go beyond a simple linear fit strategy (sec.~\ref{sec:beyond}).

\subsection{MIMAC track measurement strategy}
\label{sec:mimac}
The MIMAC prototype is the elementary chamber of the future large matrix. It enables the possibility to 
show the ionization and track measurement performance needed to achieve the directional detection strategy.
The primary electron-ion pairs produced by a nuclear recoil in one chamber of the matrix are detected by drifting 
the primary electrons to the grid of a bulk Micromegas \cite{Giomataris1,Giomataris:1995fq,Iguaz:2011yc} and producing the avalanche 
in a very thin gap (128 or 256$\mu$m).  The electrons move towards the grid in the drift space and are projected on the pixelized anode thus allowing to get 
information on the X and Y coordinates. A bulk Micromegas \cite{Iguaz:2011yc} with a 10 by 10 cm$^2$ active area, 
segmented in pixels with an effective pitch of 424 $\mu$m is used as 2D readout.
 In order to reconstruct the third dimension of the recoil, the Z coordinate {\it i.e.} along the drift axis, 
 a self-triggered electronics has been developed. It allows to perform the anode sampling at a frequency of 50 MHz.
This includes a dedicated 64 channels ASIC \cite{Richer:2009pi} associated to a DAQ \cite{Bourrion1,Bourrion2}. Each pixel is then 
connected either to a X or a Y channel implying that a spatial coincidence between one or several X and Y channels is required to 
get a localisation of the track on the (X,Y) plane. 
The ionization energy measurement is done by using a charge integrator connected to the grid wich is sampled at a frequency of 
50 MHz. Then, to recover the kinetic energy of the recoiling nucleus, one has to know accurately the value of the Ionization Quenching 
Factor (IQF) \cite{guillaudin}.\\
With such a measurement, the X and Y coordinates are measured on the anode, while the Z coordinate is retrieved from the 50 MHz
sampling of the anode. Hence, the track is 3D reconstructed, providing the electron drift velocity is known.

As an illustration, we present on figure \ref{fig:Observable} a measured track of a Fluorine recoil candidate at 50 keVee in a gas mixture composed of 
70\% $CF_4$ + 30\% $CHF_3$ at 50 mbar. The left and the center panel of figure \ref{fig:Observable} represents the projection of the 3D track on the (X,Z) and (Y,Z) planes
respectively, while the right panel represents the derivative of the charge preamplifier, which is related to the projection of the energy loss along the Z axis. This way, 
the MIMAC readout system is able to provide a large number of observables which are:
\begin{itemize}
\item the number of time slices, {\it i.e.} the number of spatial coincidences, $N_c$.
\item the time of charge collection $\Delta t_e$ defined as the time between the minimum and the maximum of the charge integrator response.
\item the projected length $L_p$ of the track on the anode plane.
\item the positions $X_i$ and $Y_i$ of the center of gravity of each time slice in the (X,Y) plane, where $i$ refers to the time slice numerotation, with $i \leq N_c$.
 The latters are represented as the black boxes on the left and center panel of figure \ref{fig:Observable}.
\item the width of each time slice $\Delta X_i$ and $\Delta Y_i$ along the X and Y axis.
\item If the charge integrator response is perfectly known and the rising time of the charge integration sufficiently short, one can extract the 
time profile of the collected charges $Q_j$, where $j$ refers to the timing of the charge integrator profile, with $j \leq \Delta t_e$.
\end{itemize}
Hence, the MIMAC readout provides us with a number of observables $\rm N_{obs}$ which grows with the number of spatial coincidence 
$\rm N_{c}$ and the time of the charge collection $\Delta t_e$ as: 
\begin{equation}
\rm N_{obs} = 3 + 4\times N_{c} + \Delta t_e,
\end{equation}
As an illustration, in the case of the measured Fluorine track candidate shown on 
 figure~\ref{fig:Observable}, we have $\rm N_{obs} = 3 + 4\times 8 + 26 = 61$ observables. However, it should be noticed that the track measurement shown on 
 figure~\ref{fig:Observable} has been done with a charge integrator characterized by a rising time of $\sim 200$ ns,
  hence overestimating $\Delta t_e$.\\
It is noteworthy that the choice of relevant observables is experiment-dependent. The observables above listed correspond to the MIMAC
detection strategy.  The likelihood approach proposed in this paper could be used within any set of directional observables from different
experiments \cite{white}.

 \begin{figure}[b]
\begin{center}
\includegraphics[scale=0.45,angle=0]{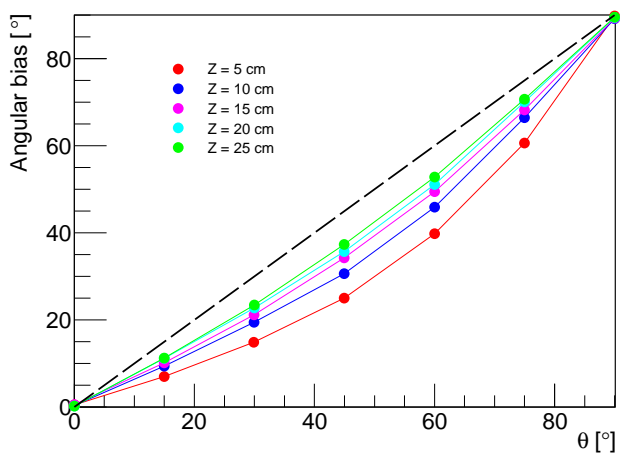}
\caption{Angular bias with respect to the original track direction as a function of the
 input value of the $\theta$ angle. This results has been obtained considering Fluorine recoils of 100 keV at \{X = 0 cm, Y = 0 cm, $\phi$ = 0$^{\circ}$\}   in a gas mixture composed of 
70\% $CF_4$ + 30\% $CHF_3$ at 50 mbar.}
\label{fig:biais}
\end{center}
\end{figure}

\subsection{The need to go beyond a simple linear fit}
\label{sec:beyond}

 A straightforward track reconstruction method consists in approximating the measured track as a straight line and to perform a linear regression of the $N_c$ centers of gravity 
 ($X_i$, $Y_i$) in three dimensions. We present on figure~\ref{fig:biais} the evolution of the angular bias with respect to the original track direction as a function of the
 input value of the $\theta$ angle. For this study the $\phi$ angle has been fixed to 0. As seen on  figure~\ref{fig:biais}, at $\theta = $50$^\circ$, the bias is about 20$^\circ$, 
 while
 it increases to  90$^\circ$ at $\theta = $90$^\circ$. It means, a track parallel to the anode is thus reconstructed, on average, as perpendicular
 track. This effect is due to electron diffusion as well as time sampling strategy.  
 Hence,  
 the estimation of the track direction   obtained with the simple
 linear regression is strongly biased. This method fails to estimate properly the direction of the recoiling nucleus in the detector. 
 Interestingly, the main explanation of this effect comes from the longitudinal diffusion. Indeed, the angular bias increases with increasing  drift distance and hence for 
 larger diffusion of the primary electrons, see fig.~\ref{fig:biais} 
 We found no bias in the estimation of $\phi$ meaning that only the estimation of the $\theta$ angle suffers from bad reconstruction. 
 Moreover, it
 should be noticed that using this straightforward track reconstruction method, it is not possible to estimate the distance between the track location in the detector volume 
 and the
 anode which is necessary to perform detector fiducialisation. 
 This study highlights the fact that is compulsory to go beyond this simple linear regression to retrieve the track properties. 
 Thus, in section \ref{sec:method}, we propose an advanced data analysis strategy based on a full likelihood approach combined with a Boosted
 Decision Tree analysis.

\section{Low energy tracks in low pressure TPC}
\label{sec:simu}
 Measuring low energy tracks, $\mathcal{O}(10)$ keV,  in low pressure (50 mbar) TPC is a difficult task requiring a high performance
readout \cite{Iguaz:2011yc} as well as dedicated electronics and DAQ \cite{Richer:2009pi,Bourrion1,Bourrion2}. Advanced data analysis 
strategy is then needed. Indeed, not only should the track be 3D measured, as described above, but a given set of parameters must be retrieved. As far as directional
detection is concerned, one must retrieve, for each track~: 
\begin{itemize}
\item the WIMP interaction point (X, Y, Z), hereafter referred to  as vertex. It is indeed referring to the beginning of the track. 
The latter is an important issue as it is related to spatial resolution and hence to the detector fiducialization used to limit contamination from surface events.
\item the initial recoil direction ($\theta, \phi$), allowing to build recoil maps needed for Dark Matter search
\cite{billard.disco,billard.profile,billard.ident,billard.exclusion}.
\item the  sense of the track. 
\end{itemize}

In the following, we give a comprehensive overview of various systematics of track measurement before presenting the simulation used in the following.

\subsection{Systematics of track measurement}
\label{sec:syste}

 \begin{figure}[t]
\includegraphics[scale=0.42,angle=0]{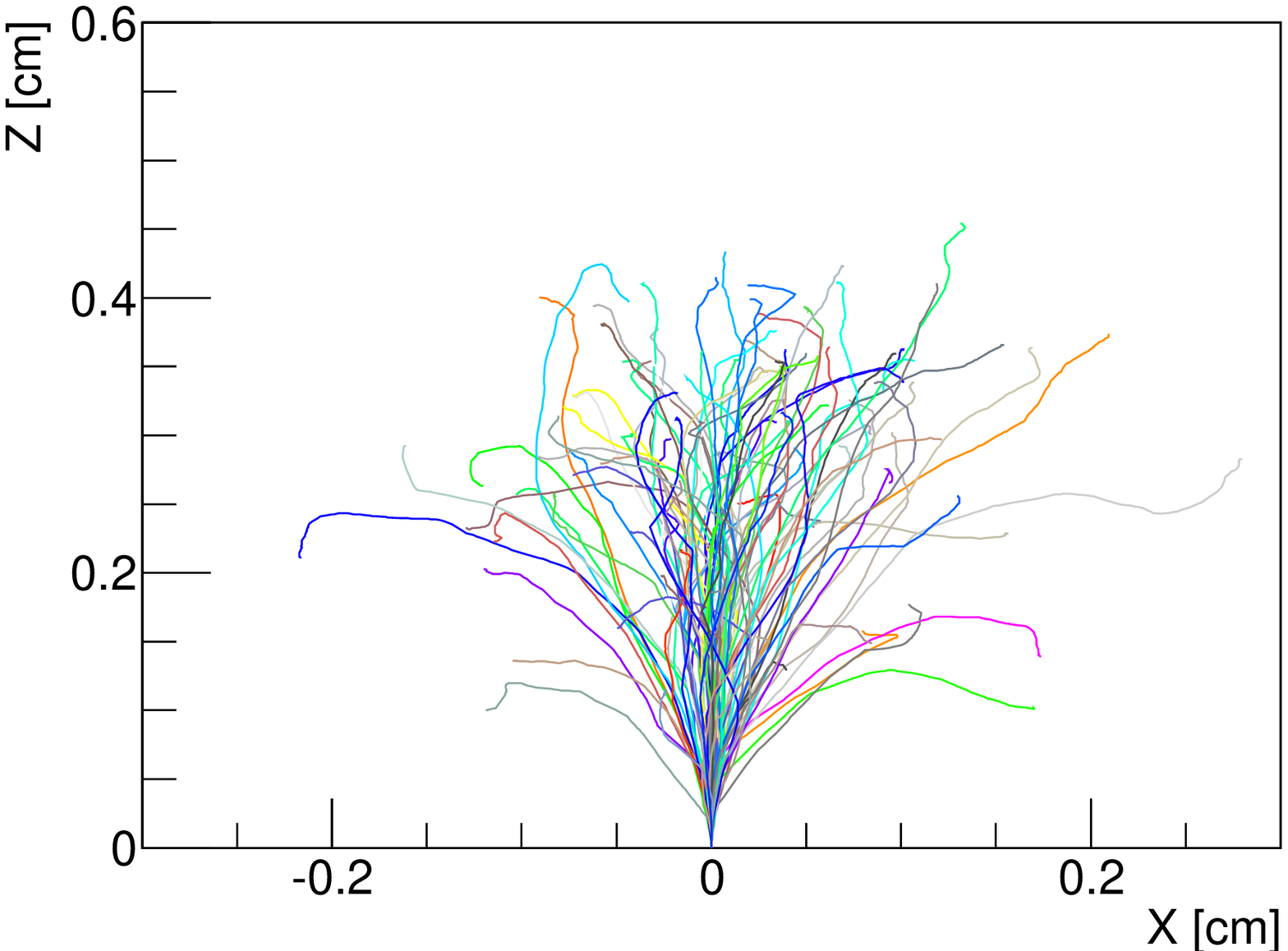}
\includegraphics[scale=0.42,angle=0]{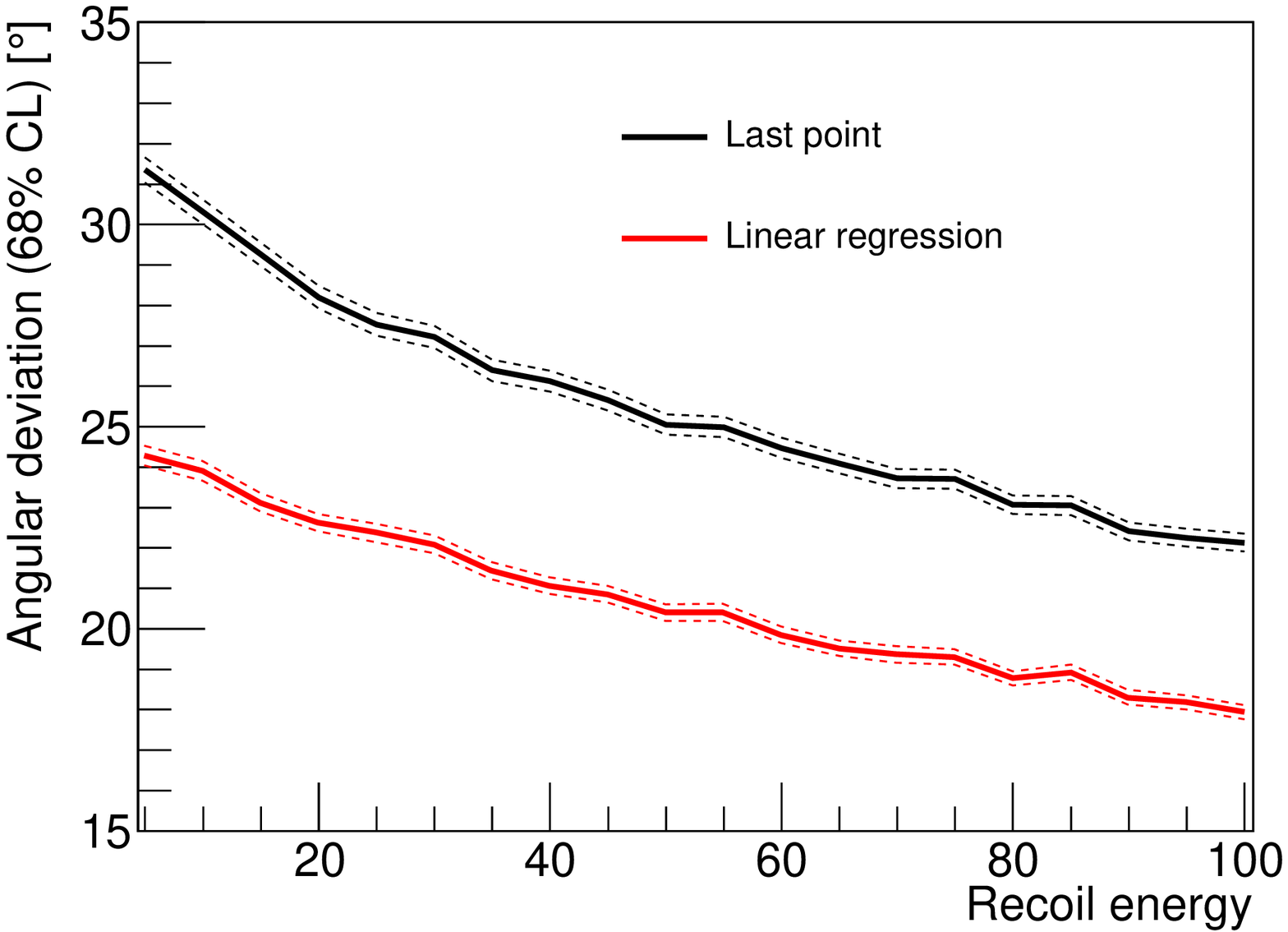}
\caption{Left: Representation in the (X,Z) plane of 100 simulated tracks of a Fluorine recoil with a kinetic energy of 100 keV in a gas mixture of 
70\%CF$_4$ + 30\%CHF$_{3}$ at 50 mbar. Right:  Angular deviation  as a function of the recoil energy, for Fluorine recoils and 
with two direction reconstruction definitions.}  
\label{fig:syste.straggling}
\end{figure}

The performance of a directional detector, in terms of resolution, is obviously strongly correlated to the track   of the target nucleus within a given gas mixture and to 
the tracking performance of the detector itself. 
In this section, we give a small review of some physical processes which contribute to the systematics associated to a track measurement in a TPC and more precisely
in the context of the MIMAC detector. Unless otherwise stated, we consider a gas mixture composed of 70\% of CF$_4$ and 30\% of CHF$_3$ at 50 mbar, but the discussion is of general interest and applies to all
directional TPC projects \cite{white}.\\ 
The angular dispersion (straggling) of low energy track in low pressure TPC is the first source of systematics that we will discuss. Indeed, the information on the initial
direction is getting lost after scattering of the recoiling nucleus with the other atoms of the gas. 
Left panel of figure \ref{fig:syste.straggling} presents 100 simulated tracks, using the SRIM software \cite{srim}, corresponding to a Fluorine recoil at 100 keV in the above mentioned gas mixture.
A large angular dispersion is observed which implies an intrinsic limitation to the angular resolution. 
We define the angular deviation as   :
$\gamma=\cos^{-1} (\hat{U}_{init}.\hat{U})$, where $\hat{U}_{init}$ and $\hat{U}$ are unitary vectors of the intial and recontructed
direction. The angular deviation $\sigma_{int}$ is then defined by :
$$
\int_0^{\sigma_{int}}f(\gamma) \ d\gamma = 0.68 
$$
Figure  \ref{fig:syste.straggling} (Right panel) presents the angular deviation  as a function of the recoil energy, for Fluorine recoils. 
The direction is reconstructed with two definitions : either the direction
between the first and the last point or a linear regression. In both cases, the intrisic angular dispersion is large and 
increasing at low energy as the straggling is getting larger.  
Typical values are : 18$^{\circ}$ at 100 keV and 25$^{\circ}$ at 10 keV. 
This strongly constrains the minimal angular resolution that a directional detector can reach.\\

Another source of systematics is the number of primary electrons generated along the track of the recoiling nucleus in the drift space of the TPC.
 For a given recoil energy $E_{r}$, the expected number of primary electrons ($\bar{N}_{elec}$) reads as   
\begin{equation}
\bar{N}_{elec} = Q\frac{E_{r}}{w}  
\label{eq:ne}
\end{equation}
with $w$ the mean energy needed to produce an electron-ion pair and Q the ionization quenching factor. The variance associated to the distribution of the number of primary
electrons is given by: $\sigma^2_{{N}_{elec}} = F\bar{N}_{elec}$, where $F$ is the Fano's factor \cite{fano} describing the spread of the distribution.  Hence, the dispersion in the number
of primary electrons will impact both the energy resolution and the spatial resolution. It is noteworthy that the ionization quenching factor is defined in such way that it
 is strictly equal to one for electrons
while for nuclear recoils it ranges between 0 and 1 depending on the nucleus, its kinetic energy ($E_r$) and the gas considered. 
This key parameter can either be estimated by simulation \cite{srim} or measured with dedicated experiments \cite{guillaudin}. 
 For instance, it is about $30\%$ for a Fluorine recoil at 20 keV, according to SRIM simulations where the ionization from secondary 
recoils have been neglected (see discussion below).\\
Once the primary electrons enter the amplification space realized with the use of the Micromegas, the number of primary electrons will be increased following an avalache
process. In a model proposed by Byrne \cite{byrne}, the gain $A$ associated to a single primary electron follows a Polya distribution as,
\begin{equation}
f(A) = \frac{(1+\Theta)^{1+\Theta}}{\Gamma(1+\Theta)}\left(\frac{A}{\bar{A}}\right)^{\Theta}\exp\left[-\frac{A}{\bar{A}}(1+\Theta)\right]
\end{equation}
where $\Theta$ is the Polya parameter.
One can find that in the case where $\bar{A}$, the mean gain, is sufficiently important, the spread of the distribution follows: $\sigma^2_{A} = \bar{A}^2/(1+\Theta)$.
 Then, the Polya parameter describes the width of the distribution which will also impact the energy resolution of the detector.\\
The amount of energy deposited in the detector in form of ionization is measured by the charge sensitive preamplifier connected to the grid.
The ionization energy is then given by the amplitude $V$ of the electronic signal. Hence, the energy resolution of the detector is
given by the dispersion $\sigma_V$ which, in the case where the electronic noise contribution to the resolution is neglected \cite{knoll}, reads:
\begin{equation}
\left(\frac{\sigma_V}{V}\right)^2 = \left(\frac{\sigma_Q}{Q}\right)^2 + \frac{w}{E_r}\left[F + (1+\Theta)^{-1}  \right]
\label{eq:resolution}
\end{equation}
where $\sigma_Q/Q$ corresponds to the fluctuations on the fraction of $E_r$ released in the ionization channel. 
As one can see from equation~\ref{eq:resolution}, the contribution to the energy resolution from the fluctuations on the number of primary electrons and on the gain of the 
Micromegas are negligible $O$(10$^{-3}$-10$^{-4}$) for recoils with $E_r \sim O(10)$ keV as $w \sim O(10)$ eV.\\

\begin{figure}[t]
\begin{center}
\includegraphics[scale=0.7,angle=0]{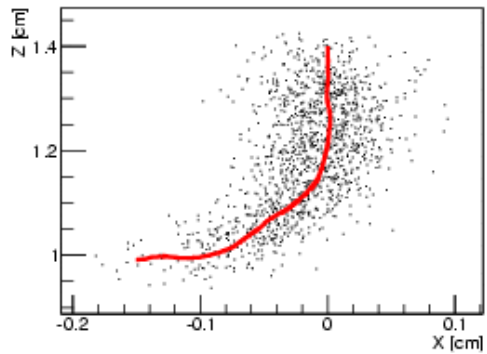}
\caption{A single Fluorine recoil track (red solid line) where we have superposed the diffused primary electrons (black dots) distribution represented in the (X,Z) plane. The
track has been generated considering a kinetic energy of 100 keV, at the location (X = 0 cm, Y = 0 cm, Z = 1.4 cm) and going toward the anode  perpendicularly to it.}
\label{fig:track.syste.cloud}
\end{center}
\end{figure}

 The electron drift within the TPC is characterized by a transverse and longitudinal diffusion. This enlarges the initial track which is described by the primary electron
 cloud.  
 During the ionization process due to the 
nuclear recoil in the gas, the kinetic energy transfered to primary electrons is of the order of 1 eV. We can then consider that primary electrons are created at
rest in the detector frame. According to kinetic theory of gas, 
the spatial distribution of the number density of electrons in co-mobile coordinates as a function of time $n(x,y,z;t)$ follows a Gaussian distribution as,
\begin{equation}
n(x,y,z;t) = \frac{n_0}{\sqrt{8\pi^3}}\frac{e^{-(x^2 + y^2)/4D_t t}}{\sqrt{4D^2_tt^2}}\frac{e^{-(z^2)/4D_l t}}{\sqrt{4D^2_l t^2}}
\label{diffusion}
\end{equation}
where $n_0$ is the initial number density of electrons and $D_l$ and $D_t$ are the longitudinal and transverse diffusion coefficient expressed in cm$^2$/s. 
The spread of the electrons along the X and Y direction,  {\it i.e.} transverse to the drift direction of electrons, is defined as $\sigma_t = \sqrt{2D_t t}$. The longitudinal spread, 
along the Z axis, reads as $\sigma_l = \sqrt{2D_l t}$. Considering the drift velocity of electrons 
$v_d$, we can rewrite $\sigma_i = \tilde{D}_i\sqrt{L}$ with $i = \{l,t\}$, $L$ the drift distance and $\tilde{D}_i$ the new expression of the diffusion coefficient expressed in
$\sqrt{\rm cm}$. 
It is worth emphasizing that the value of diffusion coefficients depends upon the unknown value of one  of the  parameters to be 
measured, namely the $Z$ coordinate of the vertex.\\
We have performed Magboltz \cite{magboltz} simulations of the diffusion coefficients as a function of the electric field for 
a  70\%$CF_4$ + 30\%$CHF_3$ at 50 mbar. In the following, we use :  $\rm \tilde{D}_t = 246.0 \mu m/\sqrt{cm}$,  
$\rm \tilde{D}_l = 278.4 \mu m/\sqrt{cm}$ and $\rm v_d = 21.4 \mu m/ns$ for  an electric field of 100 V/cm. 
The electron diffusion within the TPC volume implies a spread of the electrons of about 1 mm after a drift distance of 16 cm, which is equal to the track length of a 30 keV
Fluorine recoil in the gas considered.\\
As an illustration of the effect of electron diffusion, we show on   figure 
\ref{fig:track.syste.cloud} the representation of a single Fluorine recoil track and its diffused primary electron cloud. The Fluorine recoil has been generated considering a
kinetic energy of 100 keV, at (X = 0, Y= 0) cm,
 at 1.4 cm from the anode and perpendicular to it, going downward. The track is represented by the red solid line on the (X, Z) plane where we have superposed the
distribution of electrons generated by the nuclear recoil after their drift toward the anode. It can be
noticed that the electron diffusion will smooth the recoil track and make the detector less sensitive to the small deflections of the track.\\

Finally, even a high performance readout induces a pixelization of the track. In our case, typical 
size along X and Y is given by the pitch size whereas the Z size of the pixel is given by the the drift velocity times the sampling time.  Systematics associated with the readout is of 
course detector-dependent. In the following we list the effect associated with the MIMAC readout.\\
 Considering a grid transparency of a 100\%, we
found that at least 10 primary electrons are required per time-slice to reach a 99.9\% probability of having a spacial coincidence. This result tells us that there should
 be a quite
large difference between $N_c$ and $\Delta t_e$ as $\Delta t_e > N_c$, and that a measured track can exhibit some holes, {\it i.e.} time-slices without coincidence while electrons are still
collected at the same time. Indeed, from
Monte Carlo simulations of the detector readouts we found that a recoil track of a 20 keV Fluorine should be characterized by 30\% to 60\% of empty time slices when varying
the Z coordinate of the track from 5 cm to 25 cm. This comes from the diffusion of electrons during their drift toward the anode which will decrease the charge density per
time slice and hence decrease the probability of having a spatial concidence. This effect must be treated with caution, which is the case when using a likelihood approach
 dedicated to track reconstruction (see sec.~\ref{sec:method}).\\
We also found that, taking into account the strip size and the MIMAC trigger with the spatial concidence, the mean distance between the real center of gravity on the (X,Y)
plane and the reconstructed one is less than 1 mm even for drift distances up to 25 cm and decreases rapidly when increasing the number of electrons in the same time slice.
This result suggests that the pixel size of 424 $\mu$m is small enough as the main contribution of this spatial dispersion is due to the electron transverse diffusion. Also, we found that
the values of the observables $\Delta X_i$ and $\Delta Y_i$ increase with the drift distance meaning that they should lead to valuable information to recover the localization
of the track in the detector along the z axis.

\subsection{Track simulation}
\label{sec:Simu}
Track simulation is a key point of the analysis strategy proposed in this paper. Indeed, the method proposed (sec.~\ref{sec:method}) consists in comparing real 
track to simulated ones in order to retrieve the parameters (vertex, direction and sense) after a likelihood minimization. Hence, the track simulation must be as close as possible to real tracks in order not to
induce bias. Moreover, at the stage of validation of the method, we  reconstruct simulated tracks, in order to evaluate resolutions and bias.  
Track simulation must include all processes and systematics above described : 
\begin{itemize}
\item Nuclear recoils are propagated within the gas mixture with the SRIM software \cite{srim}, known to accurately reproduce the propagation of low energy ions in low pressure gases.
\item Primary electron generation is accounted following eq. \ref{eq:ne}. To remain conservative, the Fano factor $F$ and the Polya parameter 
$\Theta$ are taken equal to one and zero respectively, leading to Poissonian  fluctuations of the number of primary electrons created $N_{elec}$ and an exponential
distribution of the amplification gain. As   said above, the IQF is taken from SRIM neglecting the ionization from secondary recoils for computational reasons.
We found that the main contribution to the energy resolution comes from the spread of the ionization quenching. It is important to notice that neglecting the
ionization produced by secondary recoils barely modify the mean value of $Q$ at a given recoil energy but overestimates $\sigma_Q$. For example, for a 20 keV Fluorine recoil
in the considered gas mixture, we found $\sigma_Q/Q = 30\%$.
\item Electron drift properties are estimated with the Magboltz software \cite{magboltz}, allowing an accurate estimation of the drift velocity and dispersion tensor.
 In particular, both transversal and longitudinal
diffusion are accounted for. As outlined above this may lead to strong limitation in track reconstruction performance but it may also
 help in the estimation of the Z-coordinate.
\item Readout and trigger strategy are also accounted for. This is a key point as the simulated tracks must contain all systematics associated to data tracks.
 This allows us to include time sampling, pitch size, thresholds and missing coincidences. The charge
preamplifier readout is simulated by taking into account the time projection of the collected charges on the grid.
\end{itemize}
This simulation software provides us with track simulations, for various energies, drift sizes or gas properties. They are used in the following as fitting model of the likelihood minimization stage of the 3D
reconstruction method.

\section{A 2-step track reconstruction method for directional detection}
\label{sec:method}
In order to perform directional detection of Dark Matter, one must retrieve the following track parameters: vertex, initial direction and sense.  
As shown above, a straightforward track reconstruction method (a linear fit) fails at estimating accurately 
the direction of the recoiling nucleus in the detector. In particular, the estimation of the $\theta$ angle is shown to be 
strongly biased. The method proposed in this section is based on the comparison of a real track to simulated ones in order to extract the
set of parameters through a likelihood minimization. The fitting parameters associated to each tracks are : 
the recoiling energy $E_r$, the vertex $X, Y, Z$, and the direction of the track given by the
angles $(\theta, \phi)$, which leads us to 6 free parameters.\\ 
To recover accurately this set of parameters,  we introduce a 2-step track reconstruction algorithm. The first one will give a rough estimate of all the parameters (sec.~\ref{sec:first}) that will be refined during the second step which consists in  
the minimization of a likelihood function (sec.~\ref{sec:second}).


\begin{figure}[t]
\includegraphics[scale=0.42,angle=0]{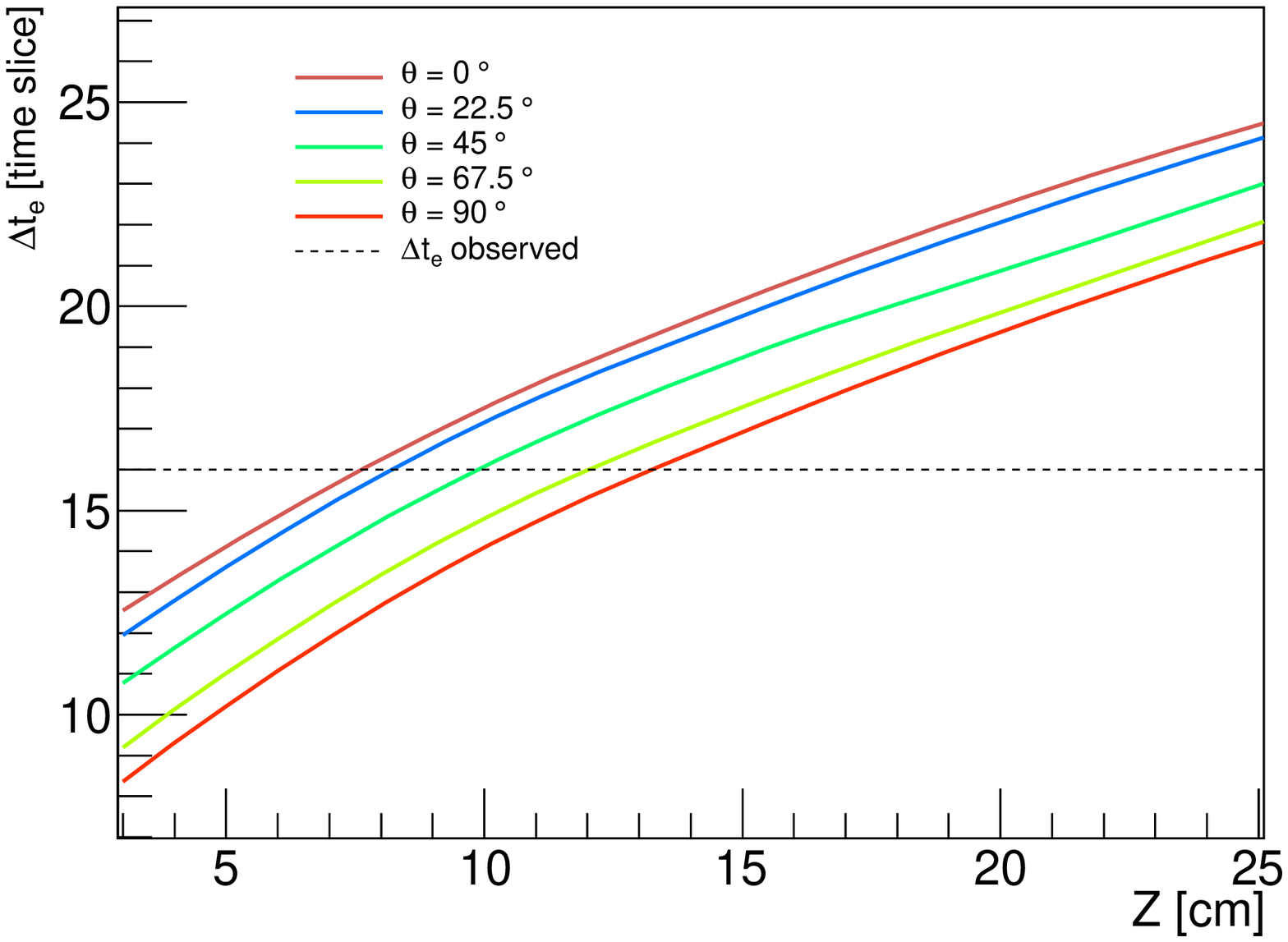}
\includegraphics[scale=0.42,angle=0]{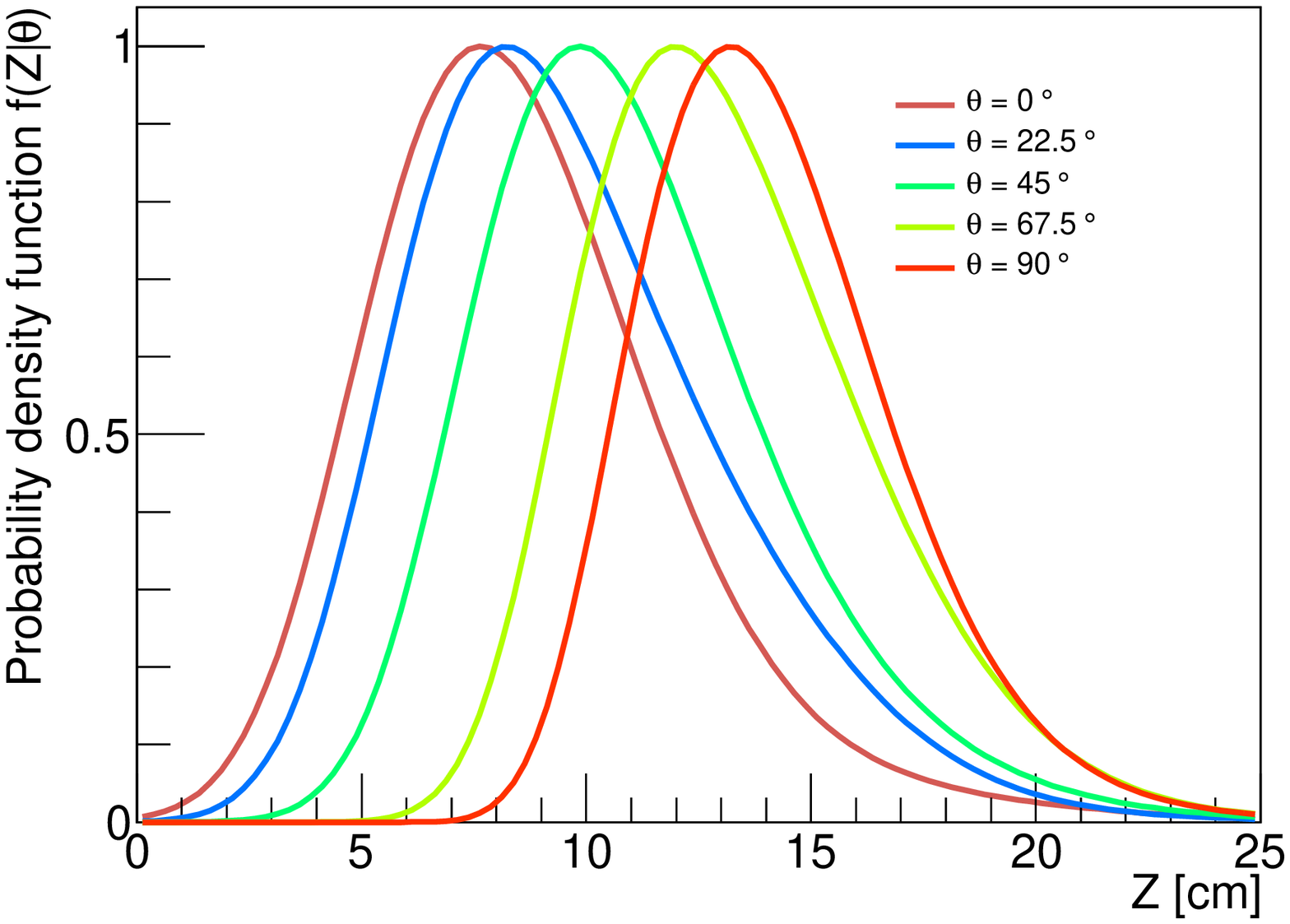}
\caption{Left: evolution of the mean value of $\Delta t_e$ as a function of $Z$ for different values of 
$\theta = \{0^{\circ},22.5^{\circ},45^{\circ},67.5^{\circ},90^{\circ}  \}$. Right: deduced probability density distribution $f(Z|\theta_i)$ for each hypothized value of
$\theta$. The two figures have been generated considering the benchmark case presented in the text.}
\label{fig:deltaE}
\end{figure}

 \begin{figure}[t]
\includegraphics[scale=0.42,angle=0]{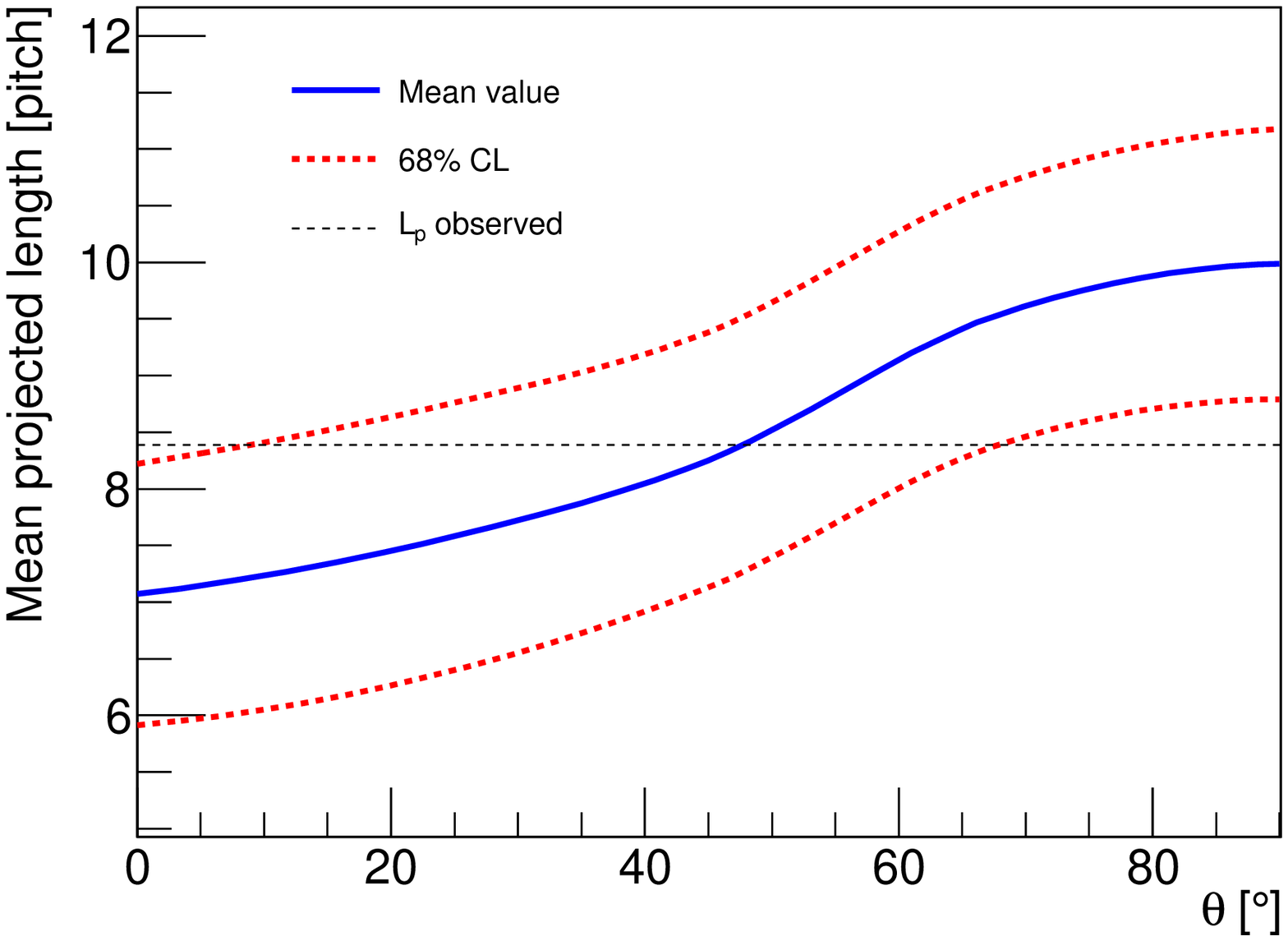}
\includegraphics[scale=0.42,angle=0]{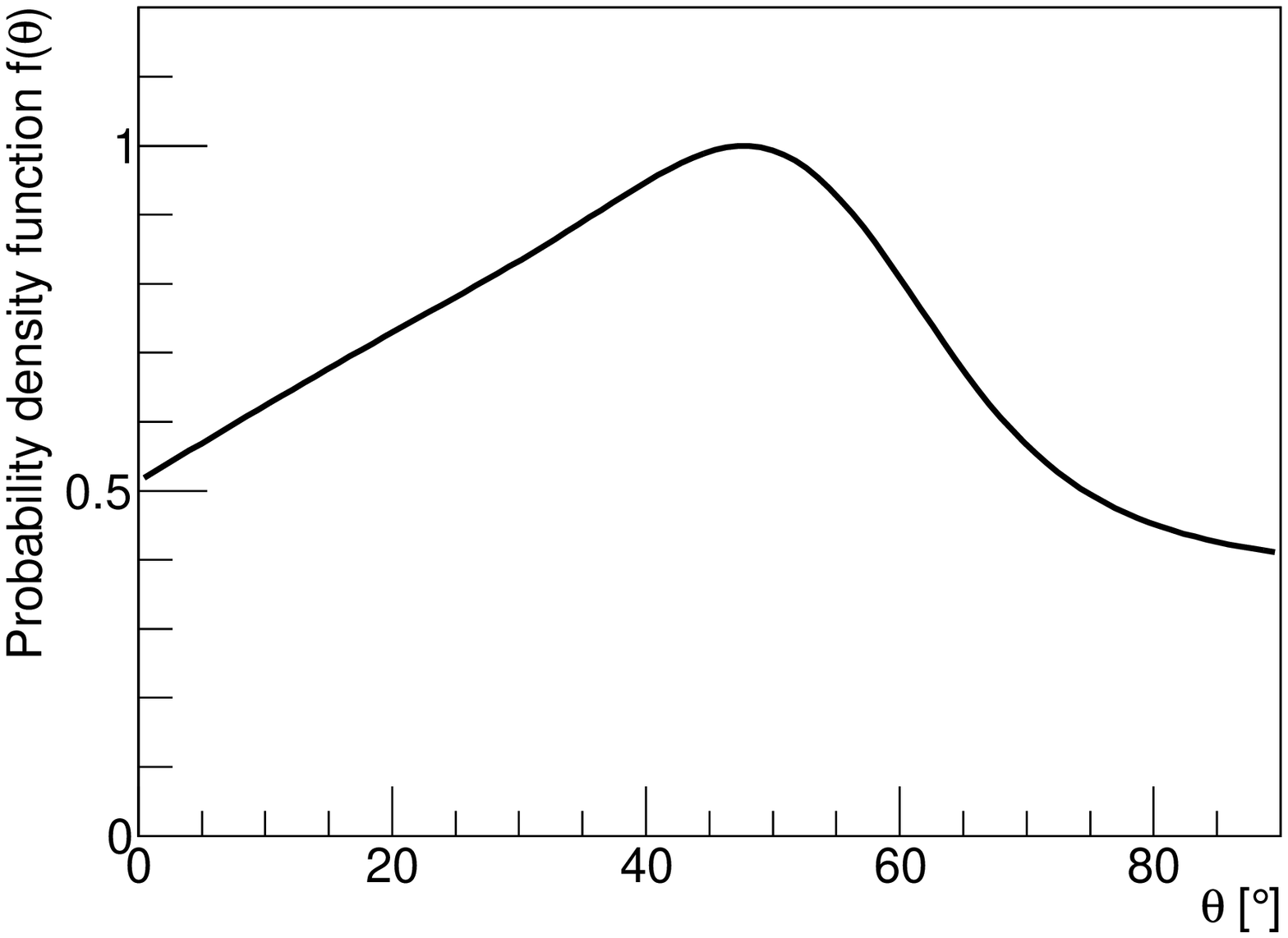}
\caption{Left: distribution of the projected length $L_p$ as a function of $\theta$. The solid line refers to the mean value of $L_p$ while the two dotted lines correspond the
68\% C.L interval. Right: deduced probability density distribution $f(\theta)$. The two figures have been generated considering the benchmark case presented in the text.}
\label{fig:projected}
\end{figure}
\subsection{First step}
\label{sec:first}
A straightforward estimation of the recoil energy $E_r$ is given by the ionization energy measurement done by the preamplifier directly connected to the grid of the MIMAC
detector. Indeed, the measurement of the ionization energy combined with the knowledge of the mean Ionization Quenching Factor (IQF) allows to recover the most probable recoil
energy corresponding to the track. Using the track measured by the MIMAC detector, it is also possible to recover an accurate estimation of the localization of the
track on the $(X, Y)$ plane and a main direction in this plane given by the angle $\phi$. This is the linear fit described in sec. \ref{sec:beyond}. However, as shown before,
this method is not able to recover an accurate estimation of the $\theta$ angle and is not sensitive to the $Z$ coordinate of the recoil track in the detector volume.\\

To allow us to have a rough estimate of $\theta$ and $Z$, we developed the following iterative procedure. For the sake of concretness, we
consider hereafter the case of a measured track having the 
following characteristics: \{$E_r$ = 100 keV, X = 0 cm, Y = 0 cm, Z = 10 cm, $\theta$ = 45$^{\circ}$, $\phi$ = 0$^{\circ}$\} and going downward,
\begin{itemize}
\item We compute the distribution of $\Delta t_e$ as a function of $Z$ for $N_{\theta}$ values of $\theta$ by simulating a large number of
tracks ($\sim 1000$) considering the
estimation of $E_r, X, Y$ and $\phi$ from the simple linear fit. On left panel of figure~\ref{fig:deltaE} is presented the evolution of the mean value of $\Delta t_e$ as a
function of $Z$ considering $\theta = \{0^{\circ},22.5^{\circ},45^{\circ},67.5^{\circ},90^{\circ}  \}$. As one can see, $\Delta t_e$ increases with $Z$ as expected from the
primary electron diffusion and obviously decreases with $\theta$.
\item From the observed value of $\Delta t_e$ we can then deduce the $N_{\theta}$ probability density functions $f(Z|\theta_i)$
 of $Z$ for the different hypothesis on the value of $\theta$. The distributions are presented on right panel of figure~\ref{fig:deltaE}.
 We can appreciate that there is
 indeed a strong correlation between $\theta$ and $Z$ and that the maximum of probability for $Z$ is between $Z$ = 8 cm and $Z$ = 14 cm depending on the considered value of
 $\theta$. Hence, one can see that the measurement of both $\Delta t_e$ and $E_r$ is able to constrain, although weakly, the Z-coordinate of the track in the detector volume.
\item We then simulate a large number of tracks following the $N_{\theta}$ different distributions $f(Z|\theta_i)$ to discriminate which value of $\theta$ is the most 
likely to correspond to the measured track.
\item From the above simulations, we use the mean projected length ($L_p$) on the $(X, Y)$ plane as a discriminant observable. Indeed, as it is shown 
on the left panel of
figure~\ref{fig:projected}, we can appreciate the evolution of $L_p$ as a function of $\theta$. The solid line corresponds to the mean value of $L_p$ while the dotted lines
refer to the 68\% C.L. interval.
\item Considering the observed value of $L_p$, one can deduce the probability density function $f(\theta)$ of the $\theta$ parameter marginalised over the $Z$ parameter.
Right panel of figure~\ref{fig:projected} presents the deduced $f(\theta)$ distribution from which we can find the estimator of $\theta$ as being the maximum of the
distribution and found to be $\theta = 47^{\circ}$ in this case.
\item Finally, we simulate one last time a large number of tracks considering the distribution $f(\theta)$ and for different values of $Z$. From the observed value of 
$\Delta t_e$ we can deduce the probability density function $f(Z)$ marginalised over $\theta$ and obtain an estimator of $Z$ defined as the maximum of $f(Z)$.
\end{itemize}
The interest of using the $L_p$ and  $\Delta t_e$ as discriminants comes from the fact that they monotonically evolves with the parameters of interest $Z$ and $\theta$.
Following this procedure, we found unbiased estimators of the $Z$ and $\theta$ parameters. This way, we are able to start the minimization of the likelihood function with a
consistent starting point $\{\hat{E_r}, \hat{X} , \hat{Y}, \hat{Z}, \hat{\theta} , \hat{\phi}\}$ estimated with the help of the first step reconstruction algorithm.\\

\subsection{Second step : a  likelihood track reconstruction}
\label{sec:second}
\begin{figure*}[t]
\begin{center}
\includegraphics[scale=0.7,angle=0]{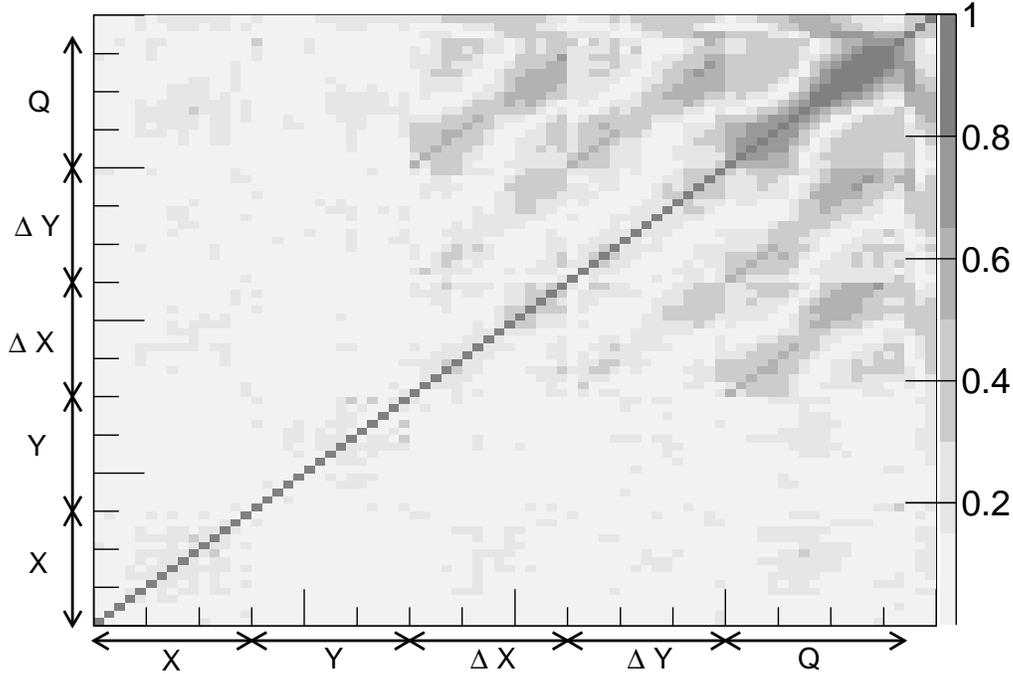}
\caption{Correlation matrix between the different observables $O^{k}$ corresponding to the case of a 50 keV Fluorine recoil located at the center of the detector going upward
in the direction $(\theta = 45^{\circ}, \phi = 45^{\circ})$ at 15 cm from the anode. The various observables are indicated on the plot : 
$ O^k= \{X_i, Y_i,\Delta X_i,\Delta Y_i,Q_j, \Delta t_e,N_c ,L_p \}$. The correlation amongst different observables is low for 
most parameters, except the correlation  between the collected charges per time slice $Q_j$ and the width of each time slice 
($\Delta X_i$ and $\Delta Y_i$).}  
\label{fig:matrix}
\end{center}
\end{figure*}

\begin{figure*}[t]
\includegraphics[scale=0.7,angle=0]{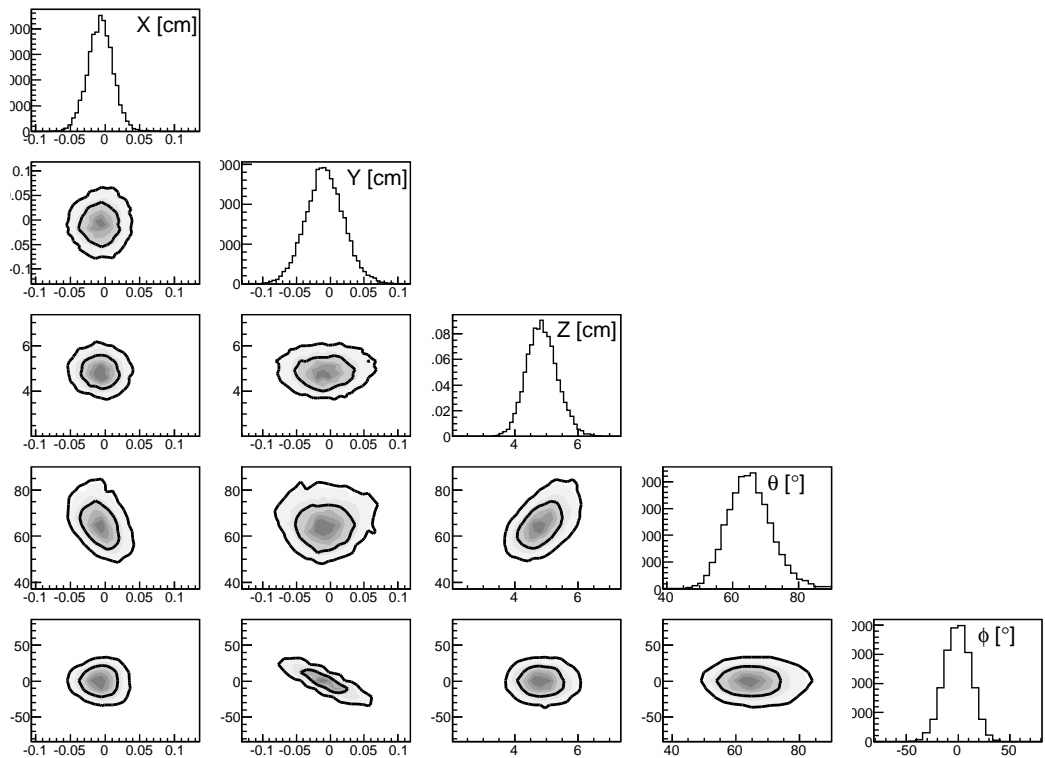}
\caption{Marginalized distributions (diagonal) ad 2D correlations (off-diagonal) plots of the likelihood function associated to a fluorine recoil of 100 keV generated at
the position \{X=0,Y=0,Z=5cm\}, and in the direction $(\theta= 45^{\circ},\phi = 0^{\circ})$.}  
\label{fig:MCMCFluor}
\end{figure*}

To go further and to improve the track reconstruction, we
developed a likelihood analysis. The idea is to compare real track to simulated ones, which thus stand as the {\em fitting model}, 
in order to extract an accurate estimate of the set of observables through a likelihood minimization. The likelihood function is defined as follows :
\begin{eqnarray}
\mathscr{L}(\vec{P}|\vec{D}) & =  \times\prod_{i=1}^{N_c}\mathscr{L}(\vec{P}|X_i)\times\mathscr{L}(\vec{P}|Y_i)\times\mathscr{L}(\vec{P}|\Delta X_i) \times\mathscr{L}(\vec{P}|\Delta Y_i) \nonumber \\
& \times\prod_{j=1}^{\Delta t_e}\mathscr{L}(\vec{P}|Q_j) \nonumber \\
& \times\mathscr{L}(\vec{P}|\Delta t_e)\times\mathscr{L}(\vec{P}|N_c)\times\mathscr{L}(\vec{P}|L_p)
\end{eqnarray}
where the $\rm N_{obs}$ likelihood functions $\mathscr{L}_k(\vec{P}|O^k)$ are defined as,
\begin{equation}
\mathscr{L}_k(\vec{P}|O^k) = \exp\left\{-\frac{1}{2}\left(\frac{O_{\rm obs}^k - \bar{O}^k}{\sigma_{\bar{O}^k}}\right)^2  \right\}
\end{equation} 
and $O^k$ refers to the different observables: 
$$ O^k= \{X_i, Y_i,\Delta X_i,\Delta Y_i,Q_j, \Delta t_e,N_c ,L_p \}$$
with  $1 \leq k \leq {\rm N_{obs}}$.\\
For the k-th observable, $O_{\rm obs}^k$ corresponds to the observed value, $\bar{O}^k$ to the expected value 
and $\sigma_{\bar{O}^k}$ to its standard deviation. Note that both the standard
deviation and the expected value are estimated from Monte Carlo simulations of 1000 tracks generated   
at the location $\vec{P} = \{E_r,X, Y, Z, \theta, \phi\}$ in the parameter space. Then, the likelihood function assumes a gaussian distribution of the different observables, which is an approximation that we have checked to
be valid. Also, it assumes that the different observables are independent to each other which means that they necessarily are uncorrelated to each other. To check this
statement, we have computed the correlation matrix defined as follows:
\begin{equation}
\rho^{i,j} = \rho[O^i,O^j] = \frac{{\rm cov}[O^i,O^j]}{\sqrt{{\rm var}[O^i]{\rm var}[O^j]}}
\end{equation}
Figure \ref{fig:matrix} presents the absolute value of the correlation matrix 
$\rho^{i,j}$.  It can be first noticed that the correlation amongst different observables is low for most parameters. 
However, we can clearly see that the collected charges per time slice $Q_j$ are strongly correlated to the width of each time slice along the $(Ox)$ axis $\Delta
X_i$ and along the $(Oy)$ axis $\Delta Y_i$. Indeed, for a large number of charge per time slice, there is an increase in sensitivity to the tail of the transverse
 electron dispersion.  However, in the following we have neglected these correlations for the following reasons:
\begin{itemize}
\item The mean of the absolute value of the correlations is $\sim 0.15$ and 97\% are below 0.5. We found that taking into account the full covariance matrix 
induces only small deviations with respect to the case without correlations. Moreover, these corrections are lower than the 
statistical fluctuations of the likelihood function caused by the fact that $\bar{O}^k$ 
and $\sigma_{\bar{O}^k}$ are estimated via Monte Carlo simulations. Hence, neglecting 
 the correlations between the different observables appears to be a fair 
approximation.
\item the size of the covariance matrix being really large ($73\times 73$ in the case presented on fig. \ref{fig:matrix}),  we found 
some numerical instabilities when computing the inverse of the covariance matrix. 
\end{itemize}
As an illustration of the method, we show on figure \ref{fig:MCMCFluor} the likelihood function, sampled using a Markov Chain Monte Carlo (MCMC) algorithm,
associated to a fluorine recoil of 100 keV. The latter was simulated at $\{X=0,Y=0,Z=5$ cm \} in the
direction ($\theta= 45^{\circ}, \phi = 0^{\circ}$) and going upward. This way, the result shown in figure~\ref{fig:MCMCFluor} 
corresponds to the same conditions used to estimate the sense recognition efficiency, except the sense (see sec.~\ref{sec:bdt}).
As one can see on figure \ref{fig:MCMCFluor}, the five parameters are strongly 
constrained. Even the $\theta$ angle and the $Z$ position which are the most challenging parameters are fully recovered with a small shift for the $\theta$ angle due to
initial angular deviation of the reconstructed track. It is important to highlight the fact that the analysis is unbiased as shown in figure~\ref{fig:ScatterPlotUpDown}. From
figure~\ref{fig:MCMCFluor} we can appreciate the fact that, due to the values of the input parameters, the $\theta$ angle is negatively correlated with the $X$ coordinate while
it is positively correlated with the $Z$ coordinate, as already explained in the previous section. Finally, we can also notice the negative correlation between the $\phi$
angle and the $Y$ coordinate.\\
 This study,  highlights the fact that a likelihood approach to track reconstruction of a few tens of keV allows to
recover consistently the track properties. Another interest of using such a likelihood analysis is that for each reconstructed track, one can get both the best fit value and
 the error bars on each parameters, taking into account all the systematics associated to the track detection : both from the track properties and the tracking performance of
 the MIMAC detector. This likelihood definition could be adapted to other directional experiments  \cite{white} with a proper choice
 of observables.

\subsection{Sense recognition : a multivariate analysis}
\label{sec:bdt}
Sense recognition is often considered as the most challenging experimental issue for directional detection. 
The question to answer can be simply put as : is the track going downward or upward ($S$ = Down or Up) ?
To reach such ability, a given detector has to be  sensitive to tiny sources of asymmetry between the two hypotheses (Up and Down). 
 The reason why experimentalists believe that it may be possible to achieve sense
recognition relies on the fact that there is, indeed, two sources of asymmetry between the beginning and the end of the track :
\begin{itemize}
\item Shape asymmetry:  more deflections are expected at the end of the track with respect to 
its beginning. Hence, the beginning of the track should be more rectilinear than its end. Of course, due to electron diffusion and finite pitch size of the detector readout, this effect might be 
strongly smeared out. However, even if a realistic detector will not be able to measure all the track deflections, it should be at 
least sensitive to an asymmetry in the spread of the track between the beginning and the end of the track.
\item Charge asymmetry: at low energy, $\lesssim O(100)$ keV, the ionization stopping power is expected to be larger at the beginning 
of the track with respect to the end. This should
lead to an asymmetry in the charge integration readout that is a function of the projection of the collected charges 
along the Z axis. Obviously, if the track is parallel to the anode, the MIMAC detector will not be sensitive to this asymmetry. Note that
this is not the case for other directional detectors such as DM-TPC \cite{dmtpc} and Newage \cite{newage} for instance. 
Indeed, the latters are sensitive to the asymmetry in the charge distribution projected on the anode plane, implying an enhancement of the
sensivity to the sense of the track.\\
\end{itemize}

  \begin{figure}[t]
\includegraphics[scale=0.42,angle=0]{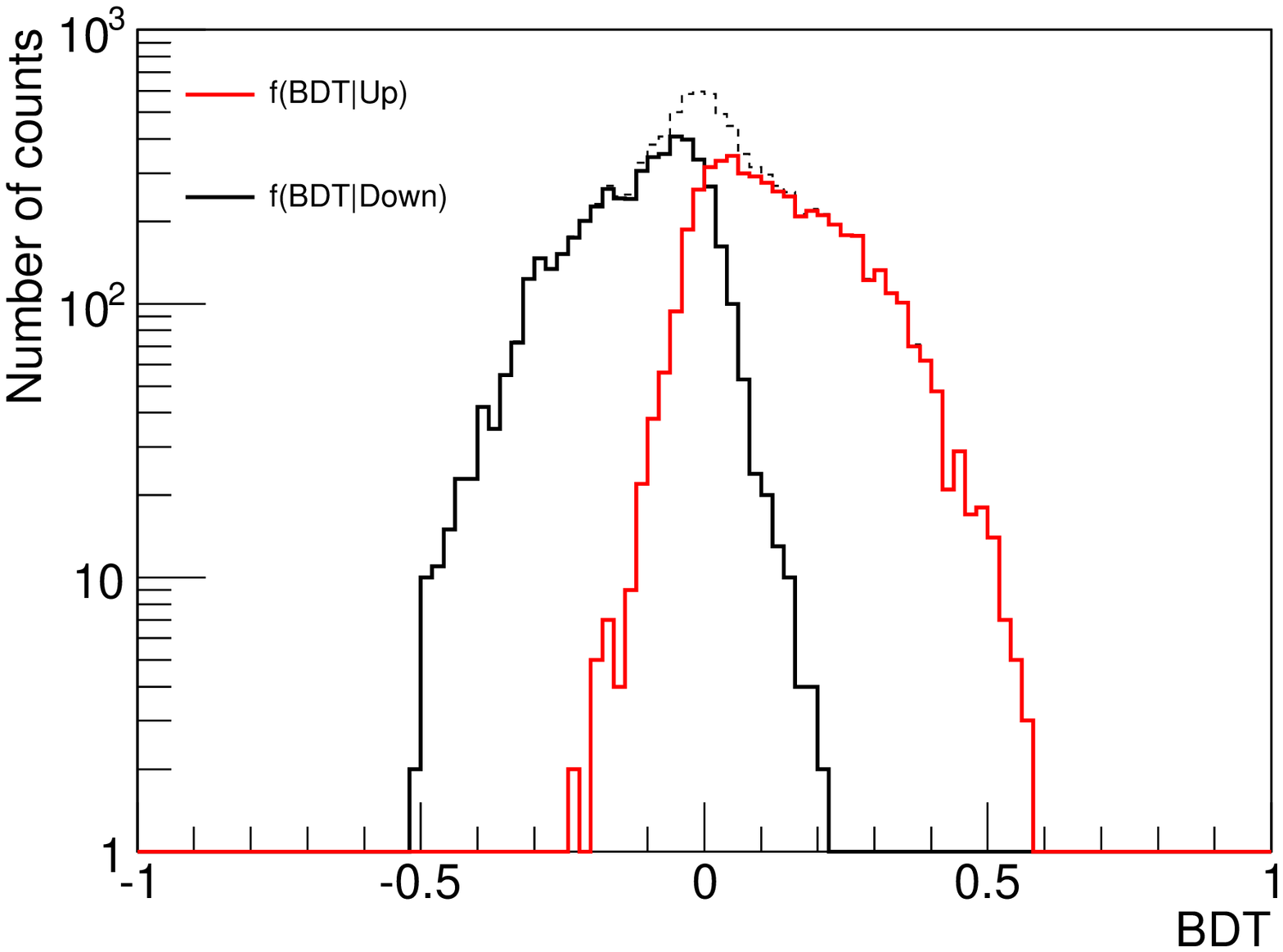}
\includegraphics[scale=0.42,angle=0]{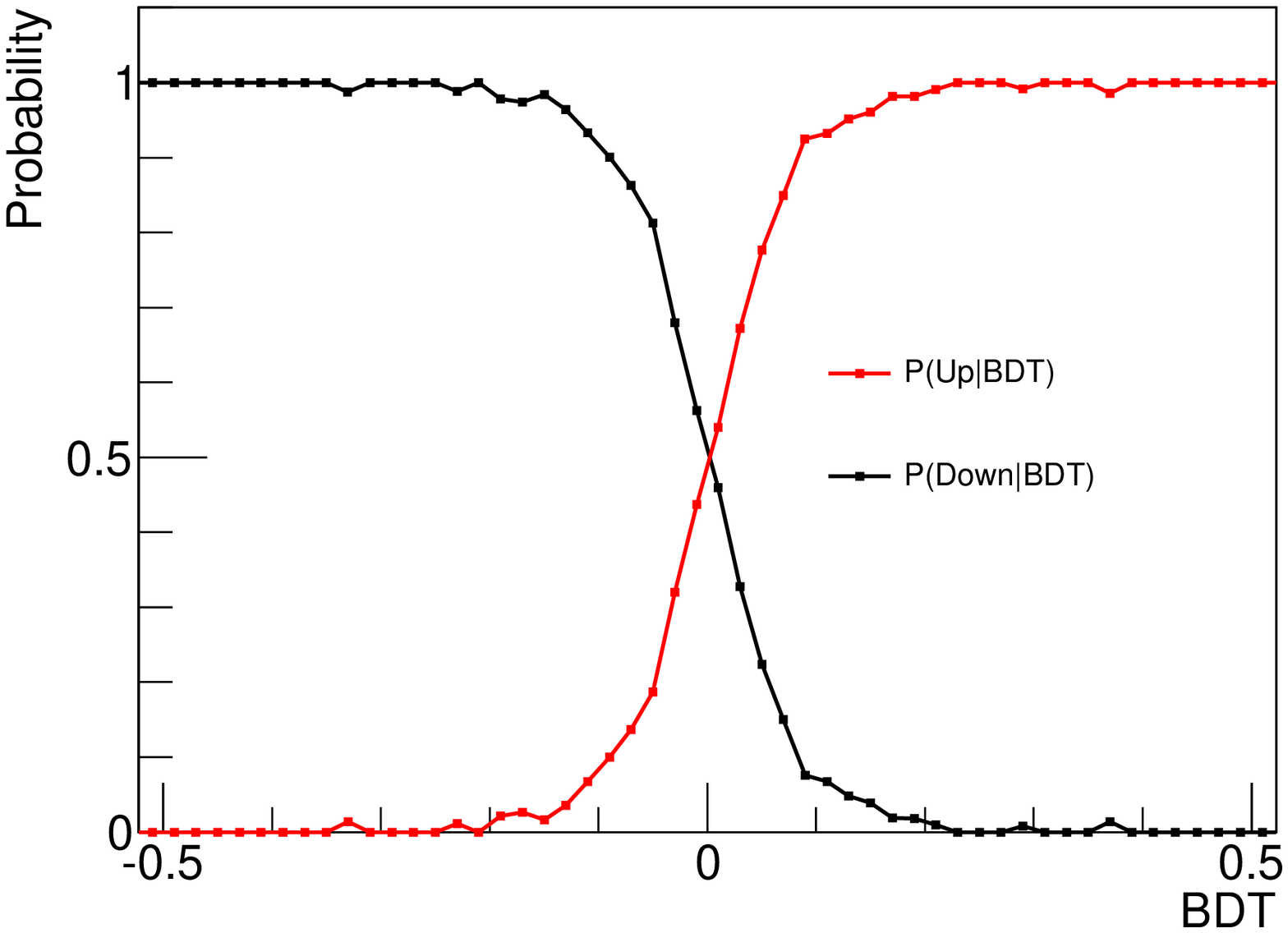}
\caption{Left: representation of $f({\rm BDT}|{\rm Up})$ (red histogram),  $f({\rm BDT}|{\rm Down})$ (black histogram) and  $f({\rm BDT}|{\rm Up + Down})$ 
(black dashed histogram) from a Boosted Decision Tree analysis.
Right : representation of  $P(U|{\rm BDT})$ (red line) and $P(D|{\rm BDT})$ (black line) as a function of BDT deduced from the result of Boosted Decision
 Tree analysis presented in the left panel.}
\label{fig:BDT}
\end{figure}

To optimize the sense recognition efficiency, we have used a high dimensional multivariate analysis, 
namely a Boosted Decision Tree \cite{tmva}. It can be seen as a data classifier, often used for signal/background discrimination. 
In our case, it is used to discriminate the Up and Down hypotheses. The principle is based on the optimisation of linear cuts 
on the different observables taken
 into account in the analysis. For this study, the observables $\tilde{O}^l$ are :
 \begin{itemize}
 \item the position (X$_i$, Y$_i$, Z$_i$) of the center  of gravity of the first and last time slice of the track,
 \item the width along the X and Y axes of the first and last time slice of the track, 
 \item the skewness factor of the  charge collection profile. 
  \end{itemize}
 Hence, we use $11$ observables in this analysis.\\ 
 The output of a Boosted Decision Tree analysis is given by the output variable ${\rm BDT}$ following,
 \begin{equation}
 {\rm BDT} =  \sum_{k=1}^{N_{tree}}\alpha_k T_k(\tilde{O}^l)
 \end{equation}
 where $\alpha_k$ refers to the normalized weight of each tree $T_k$ and $N_{tree}$ is the number of trees considered for the boosting. The main interest of boosting is that the
 misclassification rate can be considerably reduced and can even tends to zero in the case of large value of $N_{tree}$. However, in such case, there is a non negligible risk of 
 overtraining (see discussion below).\\
  By definition, we have  
 $-1 \leq {\rm BDT} \leq 1$, BDT being negative in the case of downgoing tracks and positive for upgoing one. Left panel of figure~\ref{fig:BDT} represents the final results of a Boosted
 Decision Tree analysis, associated to a Fluorine recoil of 100 keV going downward and with (X = 0 cm, Y = 0 cm, Z = 5 cm, $\theta = 45^{\circ}$, $\phi = 0^{\circ}$), given
 by the three distributions:
 $f({\rm BDT}|{\rm Up})$ (red histogram),  $f({\rm BDT}|{\rm Down})$ (black histogram) and  $f({\rm BDT}|{\rm Up + Down})$ (black dashed histogram). The Boosted Decision Tree
 analysis is based on two samples of 10,000 simulated tracks corresponding to the two hypothesis (Up and Down) deduced from the maximization of $\mathscr{L}(\vec{P}|\vec{D})$.
  From left panel of figure~\ref{fig:BDT}, 
 one can see that there is indeed 
 a discrimination between the Down and Up hypothesis suggesting that it might be possible to achieve an efficient sense recognition of observed tracks. Once the Boosted
 Decision Tree algorithm associated to a given track is built, we compute 
 ${\rm BDT_{obs}} =  f(\tilde{O}^l_{\rm obs})$ and we can then compute the probability of the track of going Upward or Downward $p(U/D|{\rm BDT_{obs}})$ using the Bayes
 theorem:
   \begin{equation}
  P(U/D|{\rm BDT}) = \frac{f({\rm BDT}|U/D)}{f({\rm BDT}|U) + f(BDT|D)}
  \end{equation}
 Right panel of figure~\ref{fig:BDT} represents $P(U|{\rm BDT})$ (red line) and $P(D|{\rm BDT})$ (black line) as a function of BDT deduced from the result of Boosted Decision
 Tree analysis presented in the left panel of figure~\ref{fig:BDT}. Hence, one can see that for $|{\rm BDT}| > 0.2$ the probability of going either upward or downward is
 $\sim 1$. The main interest of Boosted Decision Tree analysis is that we can either apply a sharp cut as if BDT positive (negative) we consider the track as going Upward
 (Downward) or just consider the relative probability and propagate it into the following Dark Matter analysis.\\
 
 A key advantage of Boosted Decision Tree analysis, in comparison with a neural network for example, is the fact that adding a new observable, even if its discrimination power
 is very weak, will never degrade its efficiency and will not change the calculation time. Then, Boosted Decision Tree can easily handle a large number of
 observables and still requires a very little calculation time. However, like any multivariate analysis algorithm, one has to check for overtraining. This check can be done
  using a Kolmogorov test statistic by comparing the distributions of the BDT value obtained on the ``training'' and the ``test'' samples which corresponds to half of the
 total number of simulated tracks of both hypothesis (Up and Down). If the two distributions are
 close to each other, according to the Kolmogorov test statistic, there is no overtraining and the Boosted Decision Tree is then consistent. In the following of the study
the Boosted Decision Tree analysis are constructed with 2000 trees and a minimum number of tracks per leaves of 50. 
The latter condition is a good protection against overtraining. Indeed, overtraining appears when the algorithm starts to learn the statistical 
fluctuations of the training sample.
Hence, by requiring at least 50 events in a leaf of any trees from the boosting prevents the algorithm to be sensitive to statistical
fluctuations and hence to be overtrained. With the parametrization presented above, we found no evidence in favor of overtraining in the following of our study.

\begin{figure*}[t]
\includegraphics[scale=0.7]{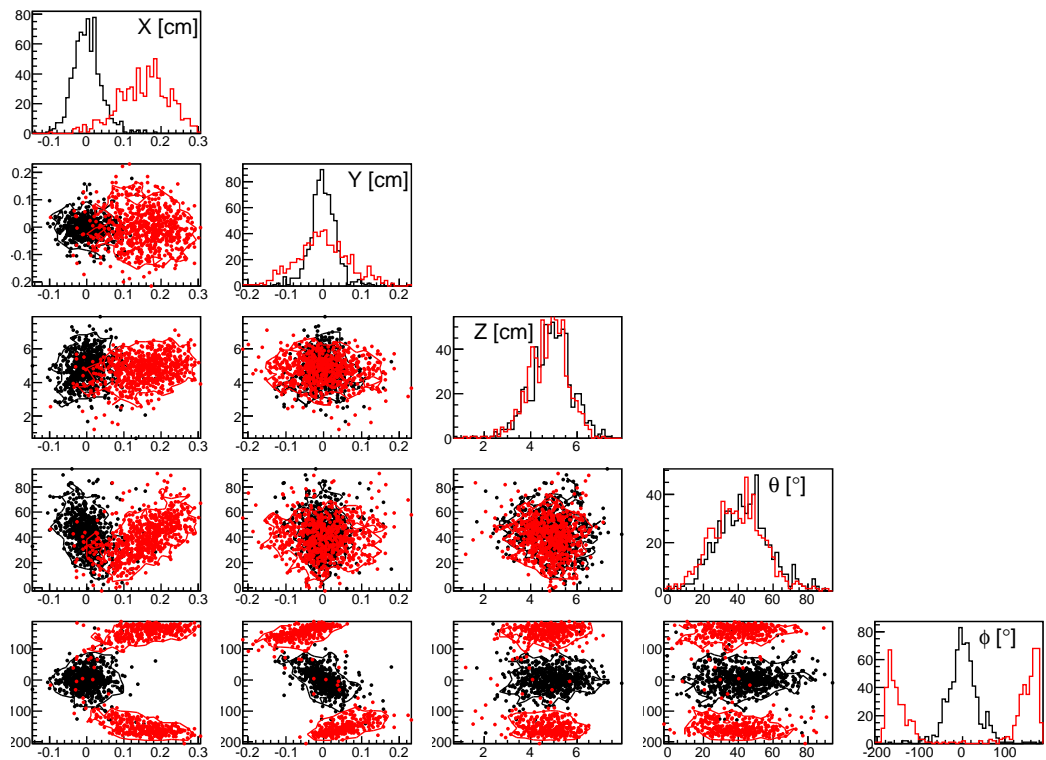}
\caption{Distribution of the maximum
 likelihood from the track analysis
of simulated data in the case of a 100 keV $^{19}$F input recoil  : $X = 0$ cm, $Y = 0$ cm, $Z = 5$ cm, $\theta = 45^{\circ}$, 
$\phi = 0^{\circ}$ and going downward. The result is presented under the two hypothesis  : Up (red histograms and red points) and Down (black histograms and 
black points).}
\label{fig:ScatterPlotUpDown}
\end{figure*}

\section{Benchmark case}
To illustrate the reconstruction algorithm, we first discuss the results corresponding to a benchmark case.
In the following, we consider the reconstruction of 800 simulated tracks corresponding to a 100 keV $^{19}$F recoil 
generated at 5 cm from the anode, at the center of the anode plane ($X = 0$ cm, $Y = 0$ cm),  
in the direction ($\theta = 45^{\circ}$, $\phi = 0^{\circ}$) and going downward. The distribution of the maximum
 likelihood under the two hypothesis  
 (Up and Down) is  presented on   
 figure~\ref{fig:ScatterPlotUpDown} in the five dimensional parameter
 space. For the sake of clarity, we have made a change of variable: $\theta \rightarrow \theta = \cos^{-1}(|\hat{t}.\hat{Z}|)$
 where $\hat{t}$ corresponds to the hypothesized direction of the track and $\hat{Z}$ to the direction of the unit vector 
 of the Z axis.\\
 We first focus on the spatial (X, Y, Z) reconstruction. As the tracks are simulated
 under the downward hypothesis, one can easily see from the marginalised 1D distribution of the five parameters 
 that there is no bias in the estimation of the different parameters. We can also extract the spatial resolution along the three axis. 
 In this case, it is about $\sim$ 350 $\mu$m along the X and Y axes ($\sigma_{x,y}$) and $\sim$ 8 mm along the Z axis ($\sigma_z$). 
 Interestingly, we can see that the 1D distribution of X, Y and Z are different under the Downward hypothesis (black histograms) with 
 respect to the Upward one (red histograms). In this  case, the track is considered back to front and the
 performance is degraded. In this case, as the simulated data track is generated with $\phi = 0^{\circ}$, 
 there is no shift in the mean value of the Y distribution under the two
 hypothesis, while a large shift is observed in the X distribution under the Upward hypothesis. One can see on fig.~\ref{fig:ScatterPlotUpDown} that 
 this shift is about 2 mm. This is what one expects, knowing that  the mean range of a 100 keV recoil,in this gas mixture and for this
 pressure, is expected to be 3 mm. The distributions are also wider under the Upward hypothesis with respect to the Downward one, 
 which is consistent with the spatial straggling of nuclear recoils presented in Sec.~\ref{sec:Simu}.\\
 Concerning the reconstruction of the direction of the track, one can see from figure~\ref{fig:ScatterPlotUpDown} that 
 the $\theta$ and $\phi$ distribution under the Downward hypothesis are consistent with the input values. It is worth emphasizing that no
 bias is observed as it was the major drawback of a linear fit method. By computing the angular dispersion distribution, we found that the associated angular resolution is $\sigma_{\gamma} =
 27^{\circ}$ at 68\% CL. We can observe a $180^{\circ}$ shift in the reconstruction of the $\phi$ angle under the Upward hypothesis, which is expected as we are considering the
 opposite direction of the track. 
 No difference between the Up and Down distributions is observed for $\theta$, as it is defined as $\theta = \cos^{-1}(|\hat{t}.\hat{Z}|)$.
 Interestingly, we can notice that the correlations between the directional parameters and the spatial parameters are of opposite sign under the Upward or Downward
 hypothesis.\\
 For this given benchmark case, we have recovered a sense recognition probability of $70\pm 1.5 \%$ where the uncertainty is given by the standard deviation of a binomial
 distribution. The latter is given by $\sqrt{\hat{p}(1-\hat{p})/N}$ where $\hat{p}$ corresponds to the observed sense recogntion probability and $N$ the number of
 reconstructed tracks.

\section{Detector performance}
\label{sec:results}
We present the expected MIMAC performance (resolutions, fiducialization, sense recognition capability) 
as a function of the energy and the Z-coordinate of the track within the  detector.  In order to estimate the angular and spatial resolutions of the MIMAC detector as well 
as the sense recognition efficiency, we used pseudo data tracks (simulated) in various configurations. This way, we have simulated 1000 tracks isotropically at fixed values of 
distance to the anode (Z) and for different values of the recoil energy ($E_r$). Each track is then reconstructed under the two hypothesis and a Boosted Decision Tree analysis 
(see sec.\ref{sec:bdt}) is used to recover the input sense.

\subsection{Spatial resolution and fiducialization}
\begin{figure}[t]
\includegraphics[scale=0.42]{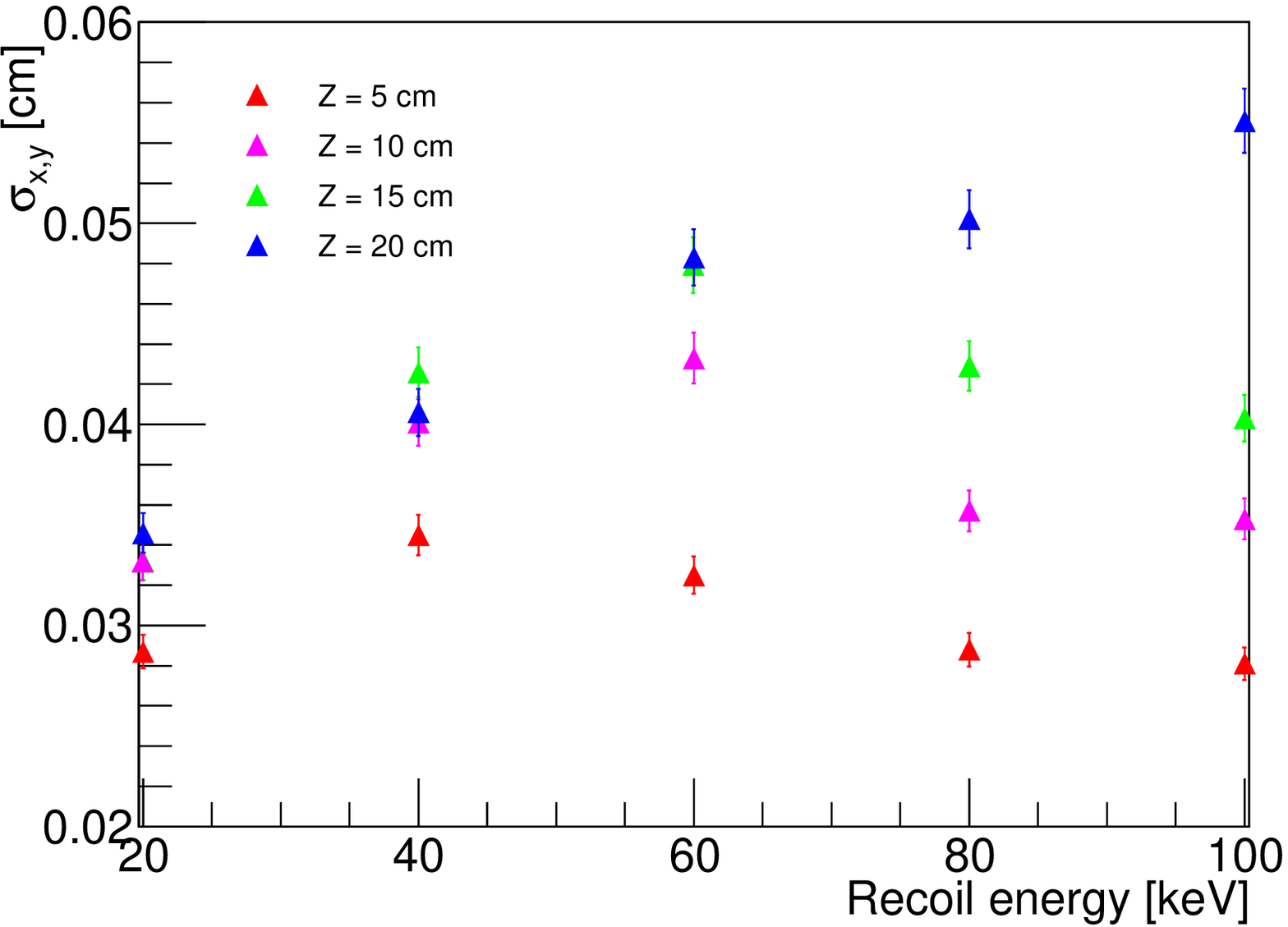}
\includegraphics[scale=0.42]{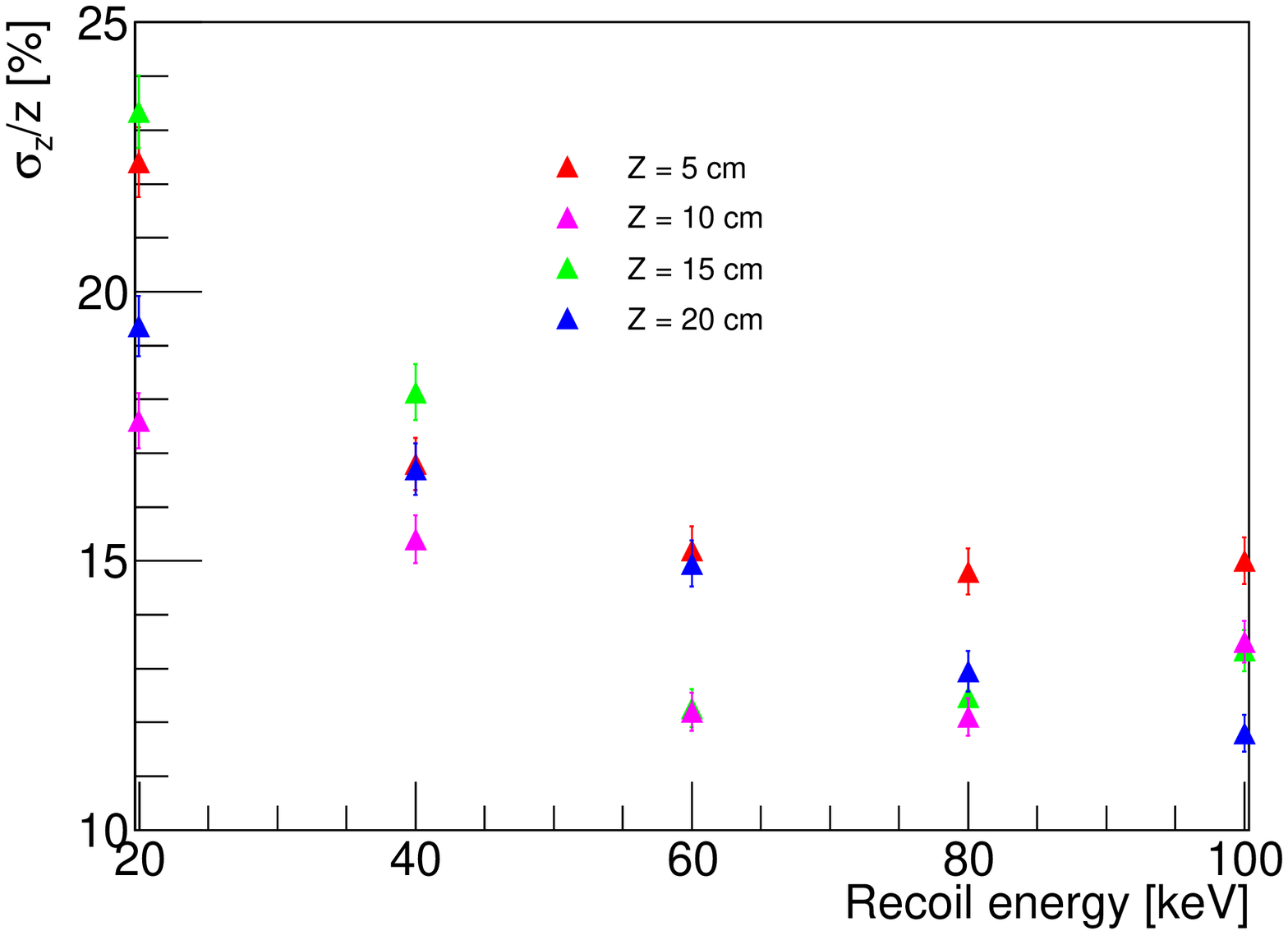}
\caption{Left : $\sigma_{x,y}$ as a function of the recoil energy. Right :   $\sigma_z/z$  as a function of the recoil energy . The result is presented 
for 4 different values of the distance between the track and the anode: Z = 5 (red), 10 (pink), 15 (green) and 20 cm (blue).}
\label{fig:spatial_resolution}
\end{figure}
To begin with, we focus on the spatial resolution which is related to the detector fiducialization.\\
Left panel of figure~\ref{fig:spatial_resolution} presents $\sigma_{x,y}$ as a function of $E_r$ and for 4 different values of the distance between the track and the 
anode: Z = 5 (red), 10 (pink), 15 (green) and 20 cm (blue). We conclude that the spatial resolution along the X and Y axes is 
ranging between 200 $\mu$m and 600 $\mu$m depending on the recoil
 energy and the distance to the anode. It is worth comparing this spatial resolution with the intrinsic resolution induced by the transversal diffusion ($\rm \tilde{D}_t = 246.0
 \mu m/\sqrt{cm}$). For instance, for a 100 keV Fluorine recoil, at 5 cm from the anode, the MIMAC resolution is $\sim 300 \ \mu$m, whereas 
 a straightforward estimation of the intrinsic resolution gives $\sim 55 \mu$m, with an average number of electrons per 
 time slice of $\sim$ 100. The fact that $\sigma_{x,y}$ increases with Z is simply due to the fact 
 that the electron diffusion coefficients evolve as the 
 square root of the  drift distance (Z). Interestingly, the evolution of $\sigma_{x,y}$ as a function of the
 recoil energy is the result of the competition between two effects. Indeed, for high recoil energy, more primary electrons are generated leading to a better
 localization of the starting point of the track. But, the track length is also increasing with energy, leading to more uncertainties on the vertex. 
 This way, increasing the recoil energy will generate more primary electrons but will also enlarge the possibility of the exact
 location of the starting point. This effect is even stronger when the electron diffusion is high, {\it i.e.} for long 
 drift distance.\\ 
 As a conclusion, a sub-mm resolution in X and Y leads to a very
 small effective volume loss when considering fiducialization (see below).\\
 The resolution along the Z axis is shown on the right panel of figure~\ref{fig:spatial_resolution}. One can see that $\sigma_z/z$ decreases with the recoil
 energy and that $\sigma_z$ increases with the drift distance (Z). As an illustration, for a 20 keV $^{19}$F recoil, $\sigma_z = 4$ cm for Z = 20 cm and $\sigma_z = 1$ cm for
 Z = 5 cm. $\sigma_z$ is larger for long drift distance due to the fact that the electron diffusion evolves as the square root of the drift distance. Indeed, the constraint
 on the $Z$ parameter evolves as $1/\sqrt{Z}$ implying a weaker constraint for long drift distances. 
  Hence, 
 the spread of the primary electron cloud, measured through the track width given by $\Delta X_i$, $\Delta Y_i$ and $\Delta t_e$, allows us to constrain the Z-coordinate of the track.  
 This constraint is even stronger for short drift distance.\\
 As a conclusion, one can see that the MIMAC experiment should reach sub-mm spatial 
 resolutions along the X and Y axis and a $\sim$ cm resolution along the Z axis.   Note that this spatial resolution could be 
 enhanced by reducing the diffusion values, by drifting ions for instance \cite{drift}.

\begin{figure}[t]
\includegraphics[scale=0.42]{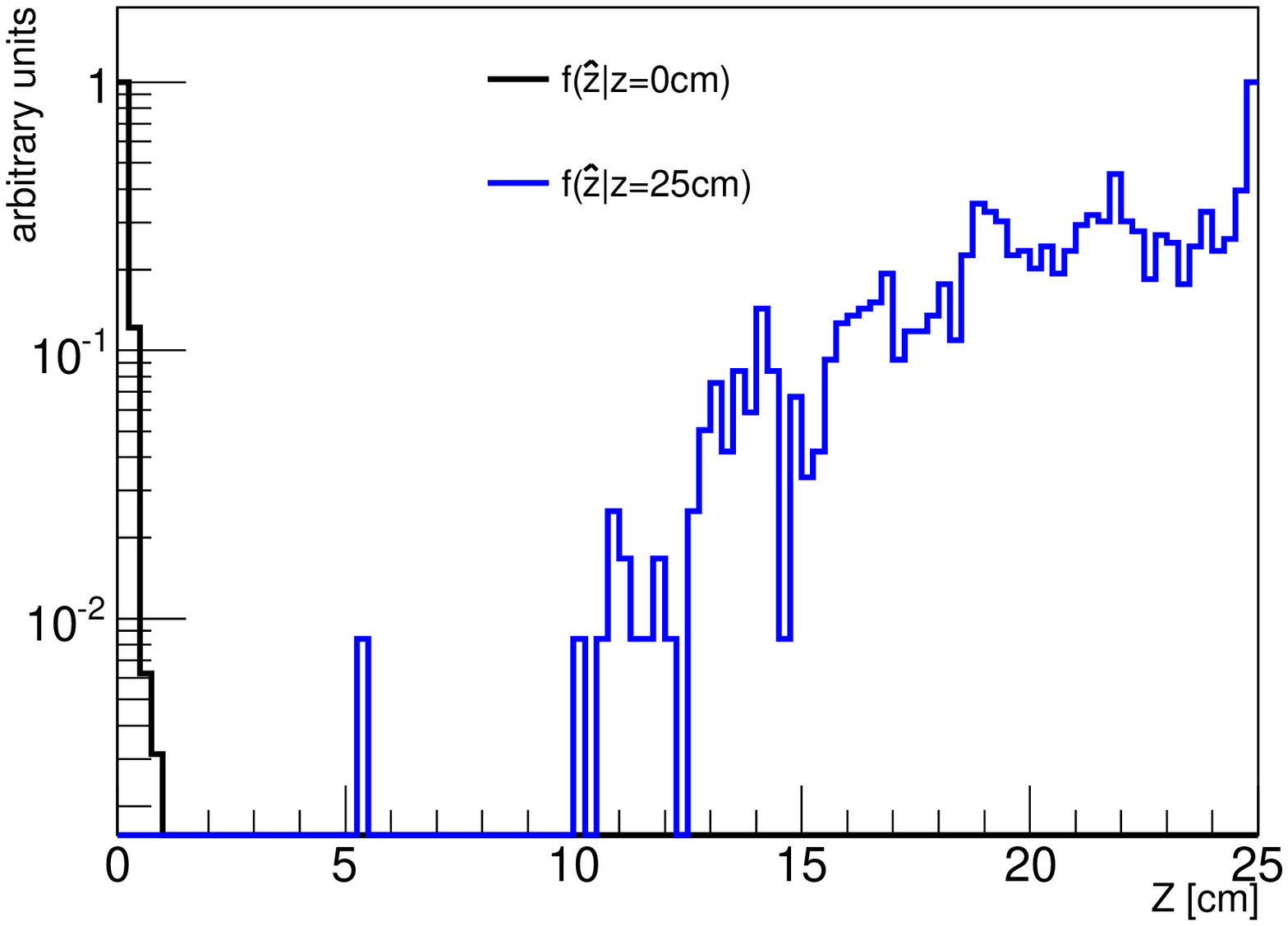}
\includegraphics[scale=0.42]{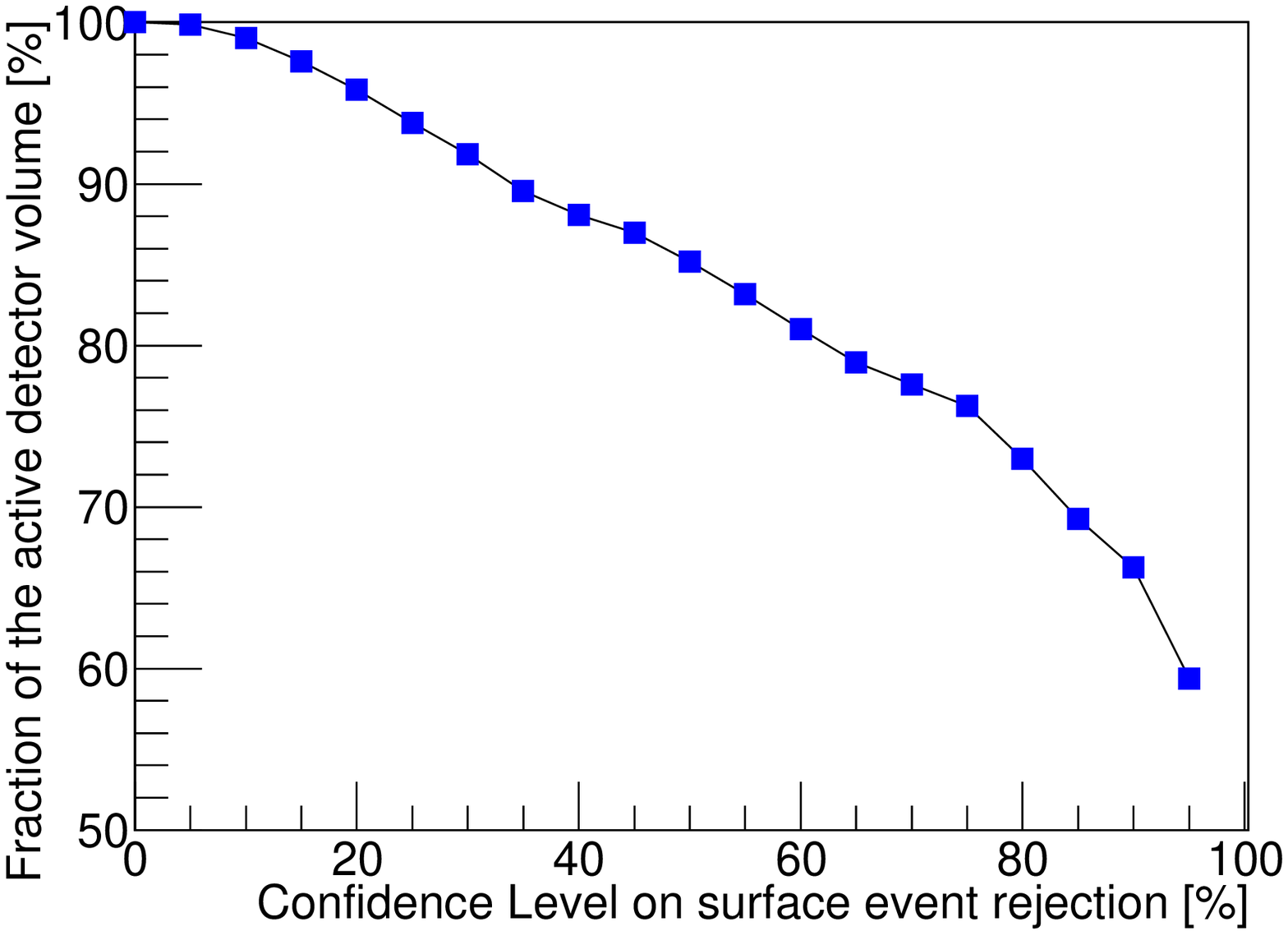}
\caption{Left : distribution of the reconstructed value of the 
Z coordinate ($\hat{Z}$). We consider the worst case respectively for the contamination from the anode (100 keV $^{19}$F recoils, in black) and 
from the cathode (20 keV $^{19}$F recoils, in blue), choosing in this case a drift distance $Z = 25$ cm. Right : fraction of the active volume as a function of $\sigma_{cut}$ the confidence
 level on surface event rejection.}
\label{fig:Fiducialisation}
\end{figure}

As above mentioned, the reconstruction of the localization of the vertex of event in the detector volume is a key issue in 
order to discriminate surface events, mostly coming from the radioactivity from the surrounding detector material. 
Then, discriminating background events will considerably reduce the background event
contamination of the data and thus improve the exclusion or discovery potential \cite{billard.profile,billard.exclusion}. 
As an illustration of surface event contamination, we present on figure \ref{fig:Fiducialisation} (left panel) the distribution of the reconstructed value of the 
Z coordinate ($\hat{Z}$) for two simulated datasets. We consider the worst case respectively for the contamination from the anode (100 keV $^{19}$F recoils, in black) and 
from the cathode (20 keV $^{19}$F recoils, in blue), choosing in this case a drift distance $Z = 25$ cm (hence the size of the chamber is 25 cm). Contrarily to what could be
assumed from right panel of figure~\ref{fig:spatial_resolution}, the 100 keV Fluorine recoil
from the anode is indeed the worst case scenario as it induces larger uncertainties on the $Z$ coordinate with repect to a 20 keV recoil. Indeed, the latter case is 
dominated by the track length at such short distance from the anode. Contamination from the anode remains 
negligeable, {\it i.e.} most anode events will be identified as such. On the contrary, a bad estimation of the vertex of a cathode event implies that it can be
counted as bulk event while it was originating from this surface. To reject surface events, a  fiducialization of the detector is proposed.
The fiducialization is computed for a given Confidence Limit (hereafter $\sigma_{cut}$). For a 25 cm drift chamber, the lower ($Z_l$) and upper ($Z_u$) limit on the Z-coordinate are given by:
\begin{equation}
\int_0^{Z_l}f(\hat{z}|z=0)d\hat{z} = \sigma_{cut} \ \ \ \ \mathnormal{and} \ \ \ \ \int_{Z_u}^{25}f(\hat{z}|z=25)d\hat{z} = \sigma_{cut}
\end{equation}
Obviously, the fiducialization should not depend on the energy of the track. We then consider the pessimistic spatial resolution on the (X,Y) plane $\sigma_{x,y} = 600 \mu$m,
 and
 along the Z axis, we have considered the $f(\hat{Z})$ distributions from the left panel of figure~\ref{fig:Fiducialisation}.
 Right panel of figure~\ref{fig:Fiducialisation} presents the fraction of active volume, {\it i.e} keeping a fiducial volume only, as a function of $\sigma_{cut}$ the confidence
 level on surface event rejection. It is then a matter of choice, balancing between purity of the events (large $\sigma_{cut}$) and size of the fiducial volume of the detector. This issue is closely related
 to radiopurity of the chosen material and Dark matter search strategy. Indeed, as outlined in  \cite{billard.profile,billard.exclusion}, 
 directional detection could accommodate to a sizeable background contamination, thanks to the 
 evaluation of the double-differential spectrum $\mathrm{d}^2R/\mathrm{d}E_R\mathrm{d}\Omega_R$. 
 This holds true even in the case of a high significance discovery of  Dark Matter. The reach of a given directional detector is thus related to the fiducial cut presented
here.\\
We emphasize that fiducialization of directional detector is possible, thanks to a good spatial resolution. The proportion of the detector volume to be removed by such analysis can be chosen with figure 
\ref{fig:Fiducialisation} (right panel),  taking into account the   radioactivity of surroundings material,

\subsection{Angular resolution}

\begin{figure}[t]
\begin{center}
\includegraphics[scale=0.5]{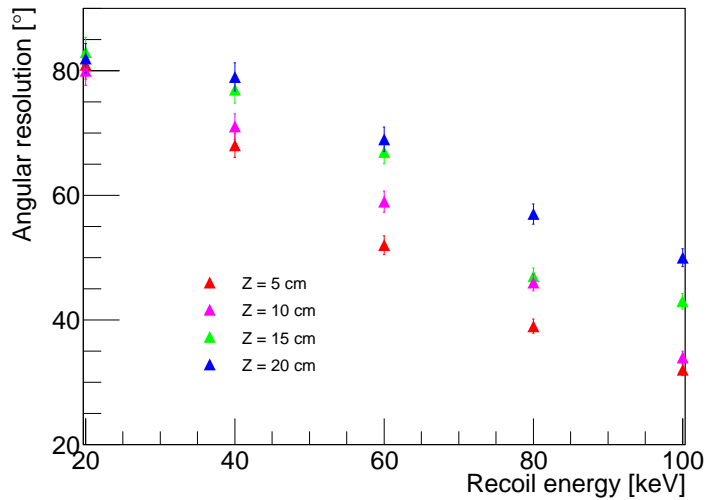}
\caption{Angular resolution as a function of the recoil energy and for 4 different
values of drift distance: Z = 5 (red), 10 (pink), 15 (green) and 20 cm (blue).}
\label{fig:angular_reso}
\end{center}
\end{figure}

The main goal of directional detection is to measure the initial direction of the nuclear recoils and the estimation of the angular resolution is then a key issue. 
It is defined as:
\begin{equation}
\int_0^{\sigma_{\gamma}} f(\gamma) d\gamma = 68\%
\end{equation}
where $\gamma$ is the angular deviation between the true direction of the pseudo data track and the reconstructed one.\\
Figure~\ref{fig:angular_reso}  presents $\sigma_{\gamma}$  as a function of the recoil energy and for 4 different
values of drift distance: Z = 5 (red), 10 (pink), 15 (green) and 20 cm (blue). It is worth emphasizing that it is the angular resolution achieved by the track
reconstruction analysis when the sense of the recoil (Up or Down) is assumed to be known. 
This way, we found that the angular resolution of the
detector should be ranging between 30$^{\circ}$ and 80$^{\circ}$, strongly depending  on the recoil energy and weakly on the drift distance. Indeed, we can see that
$\sigma_{\gamma}$ decreases when increasing the recoil energy. This effect can be easily understood. Indeed,  higher is the energy, longer the 
track and then stronger will be the constraint on its direction. It should be noticed that the angular resolution is not only due to the detector limitations, but also to the intrinsic 
 properties of low energy nuclear recoils, related to the properties of the gaz mixture. Indeed, an intrisic angular dispersion is 
 expected, due to collisions with other nuclei in the gas (see sec. \ref{sec:simu}). We have seen that a 100 keV $^{19}$F recoil has an intrisic angular dispersion of about 18$^{\circ}$ at 68\% CL. It is about 25$^{\circ}$ for a 20 keV $^{19}$F recoil. 
 Then, the value of $\sigma_{\gamma}$ observed on  figure~\ref{fig:angular_reso} is a combination of the MIMAC detector performance and of the intrinsic low energy recoil properties.\\ 
 Interestingly, at low energy, the angular
 resolution $\sigma_{\gamma}$ tends to a maximum value $\sim 90^{\circ}$ which corresponds to no directional information. Indeed, due to the fact that the likelihood
 reconstruction is based on a separation of the parameter space under two hypothesis (Up and Down), one can easily find that in that particular case, $\sigma_{\gamma} =
 90^{\circ}$ is equivalent to an isotropical guess of the track direction. Hence, the directional threshold (no directional information) lies 
 below 20 keV (recoil) in our case. Note that in the case where there is a 100\% sense recognition efficiency, the no directional 
 information regime is characterized by $\sigma_{\gamma} = 111^{\circ}$.\\
 
 As outlined in \cite{billard.profile},  angular resolution only mildly affects the sensitivity of forthcoming directional detectors. 
 In terms of disvovery potential 
 (with a 90\% probability to reach a significance greater or equal to 3$\sigma$), choosing a constant angular resolution of $60^\circ$ on the whole
 energy range, which is a fair approximation of results presented on fig~\ref{fig:angular_reso}, results in no effect at low WIMP mass with respect to the perfect angular resolution case. This is 
 due to the fact that at low WIMP mass the expected angular distribution is much more anisotropic than for heavy WIMPs. However, at high
WIMP mass, the effect of angular resolution remains small as we found a reduction in the sensitivity of about a factor of 3.3 at 1 $TeV/c^2$, with respect to the perfect angular resolution case.
The combination of this estimation of the expected angular resolution (fig.~\ref{fig:angular_reso}) together with previous results (fig. 10 in \cite{billard.profile}) allows us 
to conclude that directional detectors, as MIMAC, should reach the angular performance needed to achieve a 
high signficance discovery of Dark Matter.

\subsection{Sense recognition}

\begin{figure}[t]
\begin{center}
\includegraphics[scale=0.5]{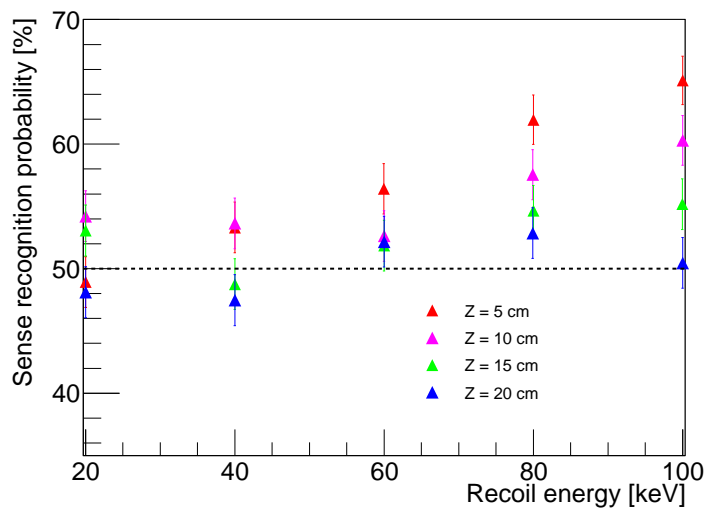}
\caption{Sense recognition efficiency as a function of the recoil energy and for different drift distances.}
\label{fig:HeadTail_eff}
\end{center}
\end{figure}

 Sense recognition has been early presented as a major issue for directional detection \cite{morgan.1,morgan.2,copi,greeen.1,vergados,green.2}.  Indeed, without sense recognition, the expected WIMP-
induced distribution becomes less anisotropic and thus gets closer to the expected background event distribution. This induces {\it a priori} a loss of discrimination
power. Even though several experimental progresses have been done  \cite{burgos,dujmic,majewski}, sense recognition remains a challenging experimental issue for directional detection of Dark Matter. 
In particular, it should still be demonstrated that sense recognition may be achieved at low recoil energy, where  most WIMP events reside. Moreover, the sense recognition efficiency must be studied.\\
Advanced data analysis is needed to recover the true sense of the recoiling nucleus and we have developed in sec.~\ref{sec:bdt}  a multivariate analysis (Boosted Decision
Trees) considering  
a large number of observables. The evolution of the sense recognition efficiency as a function of the recoil energy and for different drift distances 
is shown on figure~\ref{fig:HeadTail_eff}. Even with an advanced data analysis, applied on simulated data only,  one
   can see that the sense recognition efficiency turns out to be rather low. In the case of $^{19}$F recoils between 20 and 100 keV, it ranges between 65\% and 50\% 
   ({\it i.e.} almost no sense recognition capability).\\
   This result suggests that there is no evidence of a sense recognition capability of the detector for a  drift distance of about 20 cm while for a 5 cm drift distance, high energy nuclear recoil 
   (60-100) keV may be partially sense recognized. It is striking that sense recognition may be achieved, albeit with poor efficiency, only for events close to the anode (and at high energy). This may suggest
   either to reduce the drift size of the TPC chambers, with a huge cost when building a multi-cell detector, or to give up sense recognition. \\
The conclusion is twofold.  First, the sense recognition efficiency is shown to exhibit  a strong dependence on both the recoil energy and the drift distance. Hence, Dark matter analysis of forthcoming
directional detectors should include explicitly  this dependence in the likelihood function. This is not an easy task and this implies that the result would be strongly Monte Carlo-dependent. 
Second, the values of efficiency found in this study are rather low, suggesting that the gain  may be low when compared to the high  risk of misidentification. 
Most sudies so far \cite{billard.disco,morgan.1,morgan.2,copi,greeen.1,vergados} have used a full sense recognition on the whole energy range and on the whole detector volume, 
except \cite{green.2} that have used an energy-dependent sense recognition capability. Using an advanced data analysis strategy applied on realistic simulated data, 
we have shown that  directional detectors should not be able to reach a high sense recognition efficiency below 100
keV. We suggest not to consider this
information either to set exclusion limits or to claim a high significance discovery.\\
The effect of no sense recognition on exclusion or discovery of Dark Matter has been studied \cite{billard.profile,billard.exclusion}. As far as exclusion is concerned, the difference between 100\% sense recognition, on the whole recoil energy range and in the whole detector volume, and no sense recognition is only minor (less than a factor of three at high
background contamination). Considering a detector with no sense-recognition capability results in a loss in the discovery power of a factor of 4 at high WIMP
mass, with respect to 100\% sense recognition, and almost no effect at low WIMP mass, as outlined in  \cite{billard.profile}. A directional detector without sense recognition 
capability would still be able to achieve a 3$\sigma$ discovery in the $10^{-5} - 10^{-3}$ pb region.\\ 
As a conclusion, we have shown that sense recognition capability of directional detectors depends strongly on the recoil energy and the drift distance, with small
 efficiency values. Axial directional data (without sense recognition) seem to be a realistic goal for the future of the
 field.

\section{Conclusion}
We have proposed a new data analysis for forthcoming directional detectors. It is a  3D track reconstruction based on a 
full likelihood approach, combined with a multivariate analysis (Boosted Decision Tree). A real track is compared 
to simulated ones in order to retrieve for each track~: the WIMP interaction point (X, Y, Z), the initial recoil direction ($\theta, \phi$) and 
 the  sense of the track. This likelihood approach is applied to the MIMAC detector. 
We conclude that a good spatial resolution can be achieved, {\it
i.e.}  sub-mm in the anode plane and cm along the drift axis. This opens the possibility to perform a fiducialization of 
directional detectors. The angular resolution should range between 20$^\circ$ to 80$^\circ$ depending on the energy, which is
however enough to achieve a high significance discovery of Dark Matter. On the contrary, 
we show that sense recognition capability of directional detectors depends strongly on the recoil energy and the drift distance, 
with small efficiency values. We suggest not to consider this
information either for exclusion or discovery of Dark Mater, and then 
to focus on axial directional detectors, {\it i.e.} without sense recognition capability.\\
In a forthcoming paper we will apply this method to real tracks from measurements
with a neutron field.
 
\section*{Acknowledgments}
The authors would like to thank Cyril Grignon for fruitful helps and discussions at the early stage of this project.


\section*{References}
 

\begin{thebibliography}{10}




  
 
\bibitem{spergel}D.~N.~Spergel, 
 Phys.\ Rev.\  D {\bf 37} (1988) 1353

 
\bibitem{billard.disco}
J.~Billard, F.~Mayet, J.~F.~Macias-Perez and D.~Santos,
Phys. Lett. B {\bf 691} (2010) 156-162

\bibitem{billard.profile}J.~Billard, F.~Mayet and D.~Santos,  Phys.\ Rev.\  D {\bf 85} (2012) 035006

\bibitem{billard.ident} 
J.~Billard, F.~Mayet, D.~Santos,
Phys.\ Rev.\  D {\bf 83 } (2011)  075002
  
\bibitem{green.disco}A.~M.~Green and B.~Morgan, Phys.\ Rev.\  D {\bf 81} (2010) 061301


\bibitem{albornoz}
D. Albornoz V\'asquez, G. B\'elanger, J. Billard and F. Mayet, arXiv:1201.6150

\bibitem{billard.exclusion}
J.~Billard, F.~Mayet and D.~Santos,
 Phys.\ Rev.\  D {\bf 82} (2010) 055011 
 
 
 

 \bibitem{white}S.~Ahlen {\it et al.}, Int.\ J.\ Mod.\ Phys.\  A {\bf 25} (2010) 1



\bibitem{dmtpc}S.~Ahlen {\it et al.},
  Phys.\ Lett.\  {\bf B695 } (2011)  124-129
  
  
\bibitem{drift}
  E.~Daw {\it et al.}, Astropart.Phys. 35 (2012) 397-401 

\bibitem{d3}S.~E.~Vahsen {\it et al.},  EAS Publications Series 53 (2012) 43-50   

\bibitem{mimac}D.~Santos {\it et al.}, EAS Publications Series 53 (2012) 25-31

 
 




\bibitem{newage}
  K.~Miuchi {\it et al.},
  Phys.\ Lett.\  B {\bf 686} (2010) 11


  
 \bibitem{Giomataris1}Y.~Giomataris {\it et al.}, Nucl. Instrum. Meth. A  {\bf 560} (2006) 405   

\bibitem{Giomataris:1995fq}Y. Giomataris {\it et al.}, Nucl. Instrum. Meth. A  {\bf 376} (1996) 29 



\bibitem{Iguaz:2011yc}
  F.~J.~Iguaz {\it et al.}, JINST {\bf 6 } (2011)  P07002
  
   
  
\bibitem{Richer:2009pi}
J.~P.~Richer {\it et al.}, Nucl.\ Instrum.\ Meth.\  A {\bf 620} (2010) 470
 
\bibitem{Bourrion1}O. Bourrion {\it et al.}, Nucl. Instrum. Meth. A {\bf 662} (2010) 207  
\bibitem{Bourrion2}O. Bourrion {\it et al.},  EAS Publications Series 53 (2012) 129-136
  
 
 
\bibitem{guillaudin} O. Guillaudin {\it et al.}, EAS Publications Series 53 (2012) 119-127

 \bibitem{srim}J. F. Ziegler,  J. P. Biersack and U. Littmark U.,  Pergamon Press New York, (1985), www.srim.org.
\bibitem{fano} U. Fano, Phys. Rev. {\bf 72} (1947) 26-29

 

\bibitem{magboltz}S. F.  Biagi,  1999, Nucl. Instrum. and Meth. A, {\bf 421}, 234-240. 
\bibitem{tmva} A. Hoecker {\it et al.}, PoS ACAT 040 (2007) 

 

\bibitem{morgan.1}B. Morgan, A. M. Green and N. J. C. Spooner, Phys. Rev. D {\bf 71}  (2005) 103507 
\bibitem{morgan.2}B. Morgan and A. M. Green, Phys. Rev. D {\bf 72} (2005) 123501
\bibitem{copi}C. J. Copi, L. M. Krauss, D. Simmons-Duffin and S. R. Stroiney, Phys. Rev. D {\bf 75} (2007) 023614
\bibitem{greeen.1}A. M. Green and B. Morgan, JCAP {\bf 08} (2007)022
\bibitem{vergados}J. D. Vergados and A. Faessler, Phys. Rev. D {\bf 75} (2007) 055007
\bibitem{green.2}A. M. Green and B. Morgan, Phys. Rev. D {\bf 77} (2008) 027303 





\bibitem{burgos}S.~Burgos {\it et al.}, Astropart. Phys.  31 (2010) 261
\bibitem{dujmic}D.~Dujmic {\it et al.}, Nucl.\ Instrum.\ Meth.\  A {\bf 584} (2008) 327
\bibitem{majewski}P.~Majewski, D.~Muna, D.~P.~Snowden-Ifft and N.~J.~C.~Spooner, Astropart. Phys.  34 (2010) 284  
\bibitem{byrne}J.~Byrne, Proc. R. Soc. Edinburg Sect. {\bf A66}, 33 (1962) 
 
\bibitem{knoll}G.~F.~Knoll, Radiation Detection and Measurement

  
 
 
\end{thebibliography}
\end{document}